\documentclass[letter,traditabstract]{aa} 
\usepackage[]{graphicx}
\usepackage{txfonts}
\begin{document}
\title{The Photodetector Array Camera and Spectrometer (PACS) on the \emph{Herschel} Space 
Observatory\thanks{\emph{Herschel} is an ESA space observatory with science instruments provided by European-led Principal Investigator consortia and with important participation from NASA.}}

 \author{A.~Poglitsch\inst{1}
          \and C.~Waelkens\inst{2}
          \and N.~Geis\inst{1}
          \and H.~Feuchtgruber\inst{1}
          \and B.~Vandenbussche\inst{2}
          \and L.~Rodriguez\inst{3}
          \and O.~Krause\inst{4}
          \and E.~Renotte\inst{5}
          \and C.~van~Hoof\inst{6}
          \and P.~Saraceno\inst{7}
          \and J.~Cepa\inst{8}
          \and F.~Kerschbaum\inst{9}
          \and P.~Agn\`ese\inst{17}
          \and B.~Ali\inst{19}
          \and B.~Altieri\inst{15}
          \and P.~Andreani\inst{12,16}
          \and J.-L.~Augueres\inst{3}          
          \and Z.~Balog\inst{4}
          \and L.~Barl\inst{1} 
          \and O.H.~Bauer\inst{1}
          \and N.~Belbachir\inst{10}
          \and M.~Benedettini\inst{7}
          \and N.~Billot\inst{3}
          \and O.~Boulade\inst{3}
          \and H.~Bischof\inst{11} 
          \and J.~Blommaert\inst{2}
          \and E.~Callut\inst{5}
          \and C.~Cara\inst{3}
          \and R.~Cerulli\inst{7}
          \and D.~Cesarsky\inst{1}
          \and A.~Contursi\inst{1} 
          \and Y.~Creten\inst{6}
          \and W.~De~Meester\inst{2}
          \and V.~Doublier\inst{1}
          \and E.~Doumayrou\inst{3}
          \and L.~Duband\inst{18}
          \and K.~Exter\inst{2}
          \and R.~Genzel\inst{1}
          \and J.-M.~Gillis\inst{5}
          \and U.~Gr\"ozinger\inst{4}
          \and T.~Henning\inst{4}
          \and J.~Herreros\inst{8}
          \and R.~Huygen\inst{2}
          \and M.~Inguscio\inst{13}
          \and G.~Jakob\inst{1,12}
          \and C.~Jamar\inst{5}
          \and C.~Jean\inst{2}
          \and J.~de~Jong\inst{1}
          \and R.~Katterloher\inst{1} 
          \and C.~Kiss\inst{20}
          \and U.~Klaas\inst{4}
          \and D.~Lemke\inst{4}
          \and D.~Lutz\inst{1}
          \and S.~Madden\inst{3}
          \and B.~Marquet\inst{5}
          \and J.~Martignac\inst{3}
          \and A.~Mazy\inst{5}
          \and P.~Merken\inst{6}
          \and F.~Montfort\inst{5}
          \and L.~Morbidelli\inst{14}
          \and T.~M\"uller\inst{1}
          \and M.~Nielbock\inst{4}
          \and K.~Okumura\inst{3}
          \and R.~Orfei\inst{7}
          \and R.~Ottensamer\inst{9, 11}
          \and S.~Pezzuto\inst{7}
          \and P.~Popesso\inst{1}
          \and J.~Putzeys\inst{6}
          \and S.~Regibo\inst{2}
          \and V.~Reveret\inst{3}
          \and P.~Royer\inst{2}
          \and M.~Sauvage\inst{3}
          \and J.~Schreiber\inst{4}
          \and J.~Stegmaier\inst{4}
          \and D.~Schmitt\inst{3}
          \and J.~Schubert\inst{1}
          \and E.~Sturm\inst{1}
          \and M.~Thiel\inst{1}
          \and G.~Tofani\inst{14}
          \and R.~Vavrek\inst{15}
          \and M.~Wetzstein\inst{1}
          \and E.~Wieprecht\inst{1}
          \and E.~Wiezorrek\inst{1} \vspace{-1mm}
          }
 \institute{Max-Planck-Institut f\"ur extraterrestrische Physik,
              Giessenbachstra\ss e, 85748 Garching, Germany;
              \email{alpog@mpe.mpg.de}
              \and
              Institute of Astronomy KU Leuven,
              Celestijnenlaan 200D, 3001 Leuven, Belgium  
              \and
              Commissariat \`a l'Energie Atomique, IRFU,
              Orme des Merisiers, B\^at.~709, 91191 Gif/Yvette, France
              \and
              Max-Planck-Institut f\"ur Astronomie,
              K\"onigstuhl 17, 69117 Heidelberg, Germany
              \and
              Centre Spatial de Li\`ege, Parc Scientifique du Sart Tilman,
              Avenue du Pr\'e-Aily, 4031 Angleur-Li\`ege, Belgium
              \and
              Interuniversity Microelectronics Center,
              Kapeldreef 75, 3001 Leuven, Belgium
              \and
              Istituto di Fisica dello Spazio Interplanetario,
              Via del Fosso del Cavaliere, 00133 Roma, Italy
              \and
              Instituto de Astrofisica de Canarias, C/Via Lactea s/n,
              La Laguna, 38200 Santa Cruz de Tenerife, Spain
              \and
              Institut f\"ur Astronomie der Universit\"at Wien,
              T\"urkenschanzstra\ss e 17, 1180 Wien, Austria
              \and
              AIT Austrian Institute of Technology,
              Donau-City-Stra\ss e 1, 1220 Wien, Austria
              \and
              Institute of Computer Vision and Graphics, Graz University of Technology,
              Inffeldgasse 16/II, 8010 Graz, Austria
              \and
              European Southern Observatory,
              Karl-Schwarzschild-Str.~2, 85748 Garching, Germany
              \and
              LENS - European Laboratory for Non-Linear Spectroscopy,
              Via Nello Carrara 1, 50019 Sesto-Fiorentino (Firenze), Italy
              \and
              Osservatorio Astrofisico di Arcetri,
              Largo E. Fermi 5, 50125 Firenze, Italy
              \and
              European Space Astronomy Centre (ESAC),
              Camino bajo del Castillo, s/n, Villanueva de la Ca\~nada,
              28692 Madrid, Spain
              \and
              Osservatorio Astronomico di Trieste,
              via Tiepolo 11, 34143 Trieste, Italy
              \and
              Commissariat \`a l'Energie Atomique, LETI,
              17 rue des Martyrs, 38054 Grenoble, France
              \and
              Commissariat \`a l'Energie Atomique, INAC/SBT,
              17 rue des Martyrs, 38054 Grenoble, France
              \and
              NASA Herschel Science Center,
              Pasadena, USA
             \and
              Konkoly Observatory,
              PO Box 67, 1525 Budapest, Hungary
              }
  \date{Received 31 March 2010; accepted 28 April 2010}

 
  \abstract
{The Photodetector Array Camera and Spectrometer (PACS) is one of the three science
instruments on ESA's far infrared and submillimetre observatory. 
It employs two Ge:Ga photoconductor arrays (stressed and
unstressed) with $16\times25$ pixels, each, and two filled silicon
bolometer arrays with $16\times 32$ and $32\times 64$ pixels,
respectively, to perform integral-field spectroscopy and imaging
photometry in the 60--210\,$\mu$m wavelength regime. In photometry mode, it
simultaneously images two bands, 60--85$\,\mu$m or 85--125$\,\mu$m and
125--210$\,\mu$m, over a field of view of $\sim 1.75'\times 3.5'$, with
close to Nyquist beam sampling in each band. In spectroscopy mode, it
images a field of $47\arcsec \times 47\arcsec$, resolved into $5\times 5$
pixels, with an instantaneous spectral coverage of $\sim 1500$~km/s and
a spectral resolution of $\sim 175$~km/s. We summarise the design of the
instrument, describe observing modes, calibration, and data analysis
methods, and present our current assessment of the in-orbit performance
of the instrument based on the Performance Verification tests. PACS is
fully operational, and the achieved performance is close to or better
than the pre-launch predictions.}

 \maketitle
\section{Introduction}

The PACS instrument was designed as a general-purpose science instrument 
covering the wavelength range $\sim$60--210\,$\mu$m. It features both, a photometric 
multi-colour imaging mode, and an imaging spectrometer. 
Both instrument sections were designed with the goal of maximising the science return
of the mission, given the constraints of the \emph{Herschel} platform 
(telescope at $T \approx 85\ K$, diffraction limited for $\lambda > 80\,\mu$m, 
limited real estate on the cryostat optical bench) 
and available FIR detector technology.

\subsection {Photometer rationale}

Photometric colour diagnostics requires spectral bands with a relative bandwidth
$\Delta\lambda/\lambda < 0.5$. 
In coordination with the longer wavelength SPIRE bands, the PACS photometric 
bands have been defined as 60--85$\,\mu$m, 85--130$\,\mu$m, and 130--210$\,\mu$m, 
each spanning about half an octave in frequency.

A large fraction of the \emph{Herschel} observing time will be spent on deep and/or large
scale photometric surveys. For these, mapping efficiency is of the highest priority.
Mapping efficiency is determined by both, the field of view of the
instrument (in the diffraction-sampled case, the number of pixels) 
and the sensitivity per pixel. 
The PACS photometer was therefore designed around the largest detector 
arrays available without compromising sensitivity.

Simultaneous observation of several bands immediately multiplies 
observing efficiency.  By implementing two camera arrays, PACS can observe a 
field in two bands at a time.

Extracting very faint sources from the bright telescope background
requires means to precisely flat-field the system responsivity on
intermediate time-scales, as well as the use of spatial modulation
techniques (chopping/nodding, scan-mapping) to move the signal frequency
from ``DC'' into a domain above the 1/f ``knee'' of the system,
including - most notably - the detectors.

\subsection{Spectrometer rationale}

Key spectroscopic observations, particularly of extragalactic sources,
ask for the detection of faint
spectral lines with medium resolution ($R \sim 1500$).

The power emitted or absorbed by a single spectral line in the
far-infrared is normally several orders of magnitudes lower than the
power in the dust continuum over a typical photometric band. Sensitivity
is thus the most important parameter for optimisation; with
background-limited detector performance the best sensitivity is obtained
if the spectrometer satisfies the following conditions: The detection
bandwidth should not be greater than the resolution bandwidth, which in
turn should be matched to the line width of the source, and, the line
flux from the source must be detected with the highest possible
efficiency in terms of system transmission, spatial and spectral
multiplexing.

Again, subtraction of the high telescope background has to 
be achieved by appropriate spatial and/or spectral modulation techniques.

\section{Instrument design}

\begin{figure}[b]   
\vspace{-1mm}
   \centering
   \includegraphics[width=1.0\columnwidth]{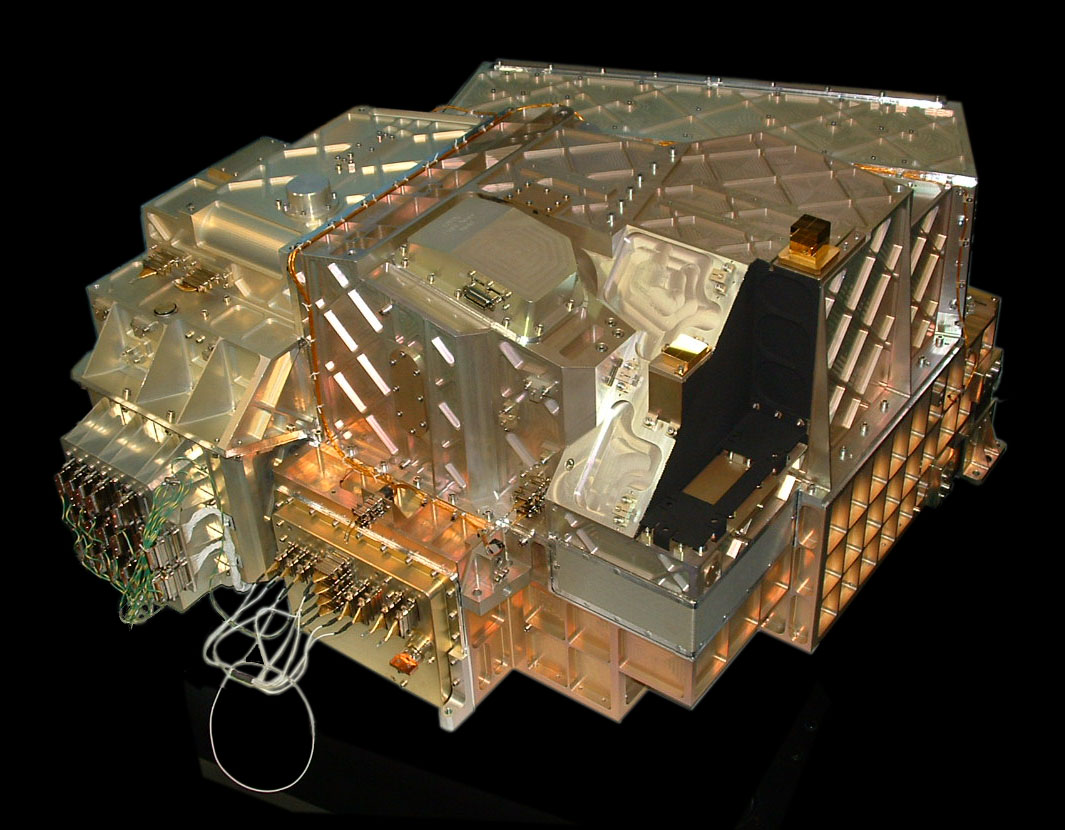}
      \caption{The PACS focal plane unit (Qualification Model)
              }
         \label{FPU}
   \end{figure}

The instrument is divided into optically well separated
compartments: A \emph{front optics section}, common to all optical
paths through the instrument, containing grey-body calibration
sources and a chopper, each at an intermediate image of the
telescope secondary. Thereafter follow the separate \emph{photometer
camera} and \emph{spectrometer} sections. The whole instrument (Fig.~\ref{FPU}) 
-- except the detectors -- is kept at the 
``Level 1'' temperature of $\sim$\,3 to 5\,K provided by the satellite.

	\subsection{Front optics}

\begin{figure*}
   \includegraphics[width=1.01\columnwidth]{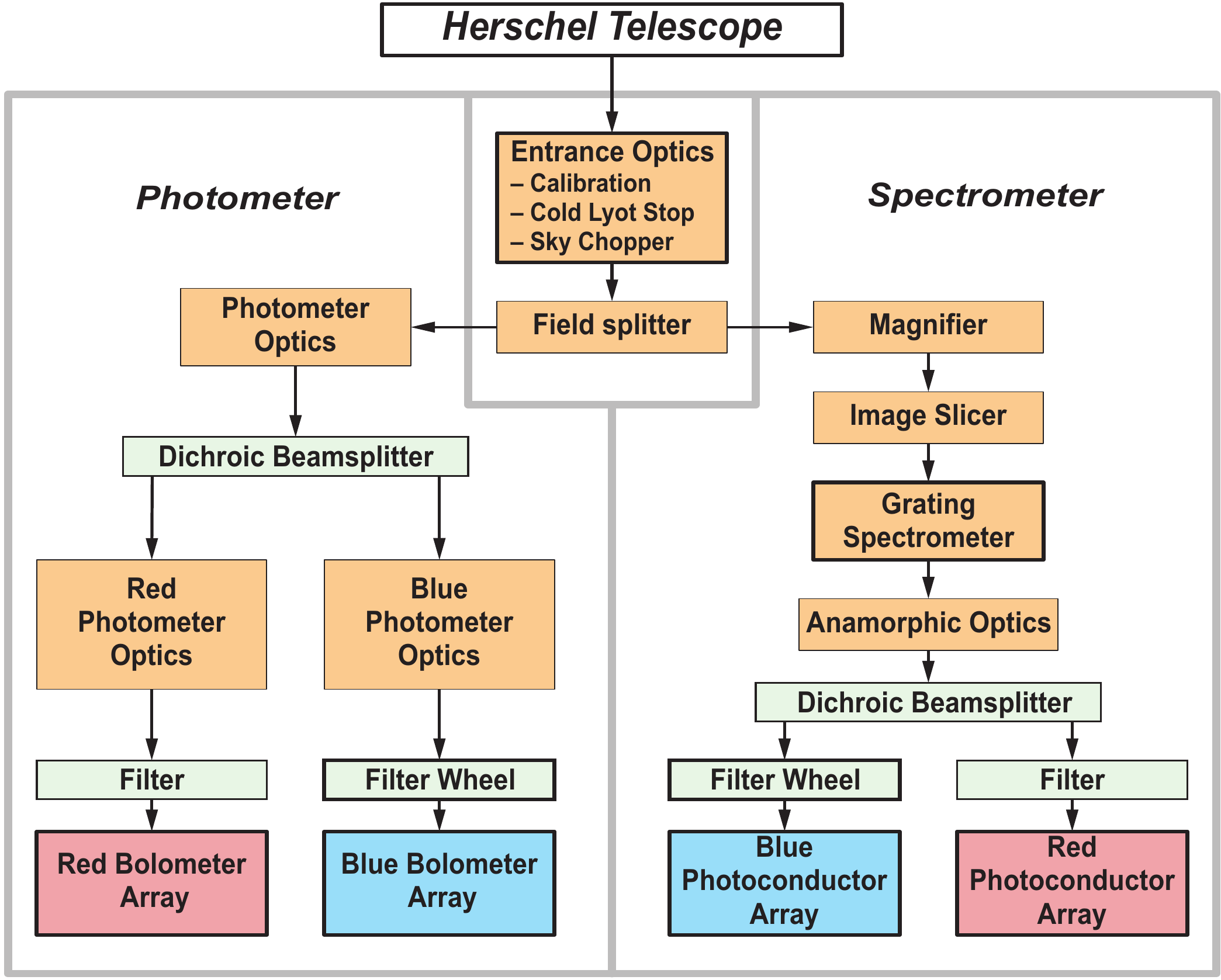}
   \hspace{1.5mm}
   \includegraphics[width=\columnwidth]{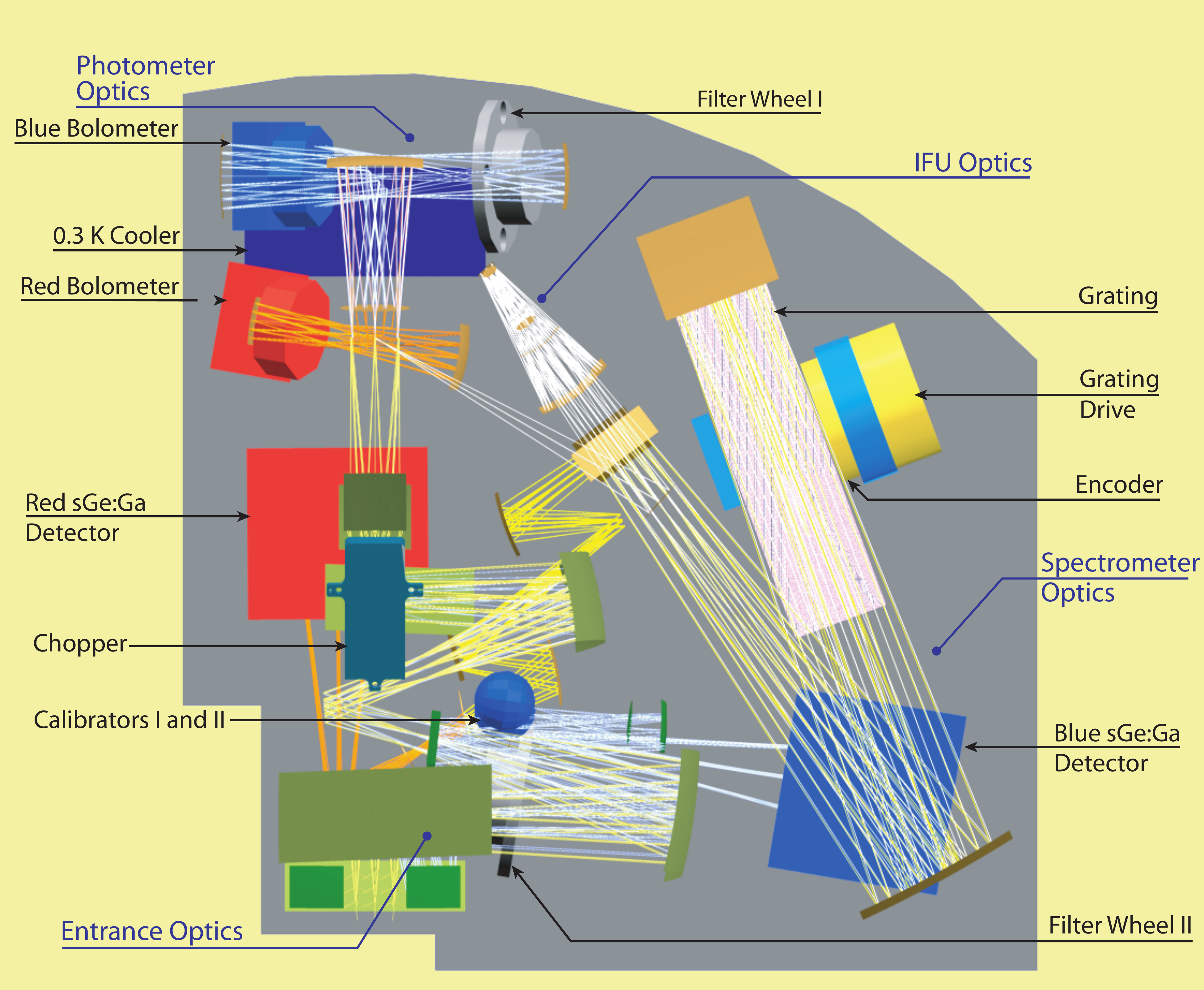}
   \caption
	{
		\emph{Left: PACS focal plane unit (FPU) functional block diagram}.
The arrows visualise the optical paths through the instrument. 
Imaging optics and filter components are shown in different colours; 
active components (mechanisms, electronics) are outlined in bold.
		\emph{Right: PACS FPU layout}.
After the common entrance optics with
calibrators and the chopper, the field is split into the
spectrometer train and the photometer trains. The two \emph{bolometer
cameras} (top) have partially separate re-imaging optics split by a dichroic
beam splitter; the short wavelength camera band is further split up by two
filters on filter wheel I. 
In the \emph{spectrometer train}, the integral field unit (image slicer, middle) first converts the
square field into an effective long slit for the Littrow-mounted
grating spectrograph (on the right). The dispersed light is distributed to the
two photoconductor arrays by a dichroic beam splitter between the 1$^{\rm st}$ and 
2$^{\rm nd}$ orders, then the 2$^{\rm nd}$ or 3$^{\rm rd}$ order for the short-wave array
is chosen by filter wheel II.
	}
     \label{fig:system}%
\end{figure*}

\begin{figure}[h]
   \centering
   \includegraphics[width=0.95\columnwidth]{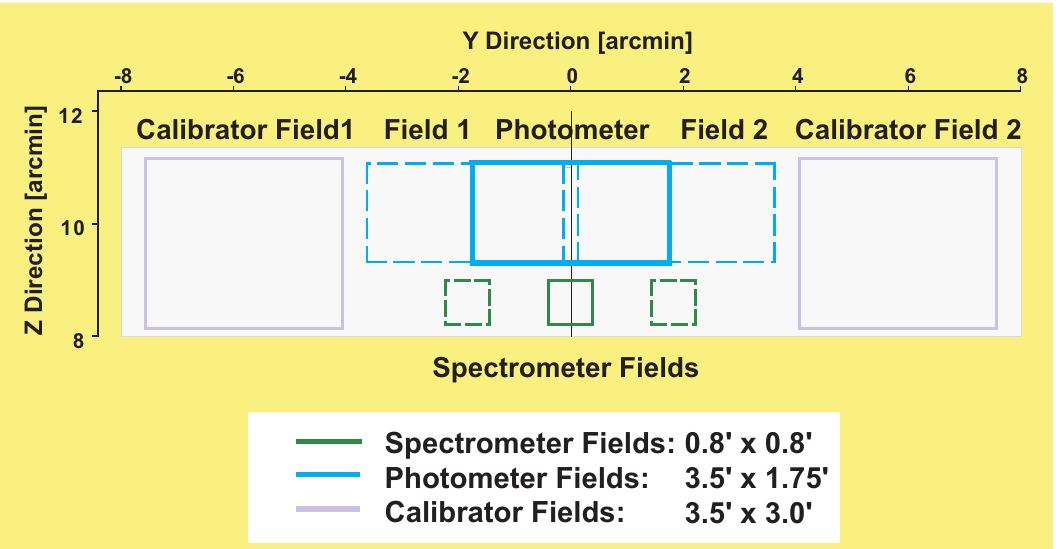}
   \caption
	{
\emph{PACS telescope focal plane usage.} Long- and short-wavelength
photometry bands are coincident. The smaller spectrometer
field of view is offset in the -Z direction. Chopping is along the Y axis 
(left-right). On both sides of the sky 
area in the focal plane the internal calibrators are reachable by the chopper. 
The maximum chopper amplitude for sky
observations (used in spectroscopy) is $\pm 3\arcmin$. 
	}
     \label{fig:focalpl}%
\end{figure}

The front optics (see Fig.~\ref{fig:system}) has several instrument wide
tasks:  It provides for an intermediate image of the telescope secondary
mirror (the entrance pupil of the telescope) with the cold Lyot stop
and the first blocking filter, common to all instrument channels. 
A further image of the pupil is reserved for the focal plane
chopper; this allows spatial chopping with as little as possible
modulation in the background received by the instrument,
and it allows the chopper -- through two field
mirrors adjacent to the field stop in the telescope focal
plane -- to switch between a (chopped) field of view on the sky
and two calibration sources (see also Fig.~\ref{fig:focalpl}).

\begin{figure}
   \includegraphics[width=\columnwidth]{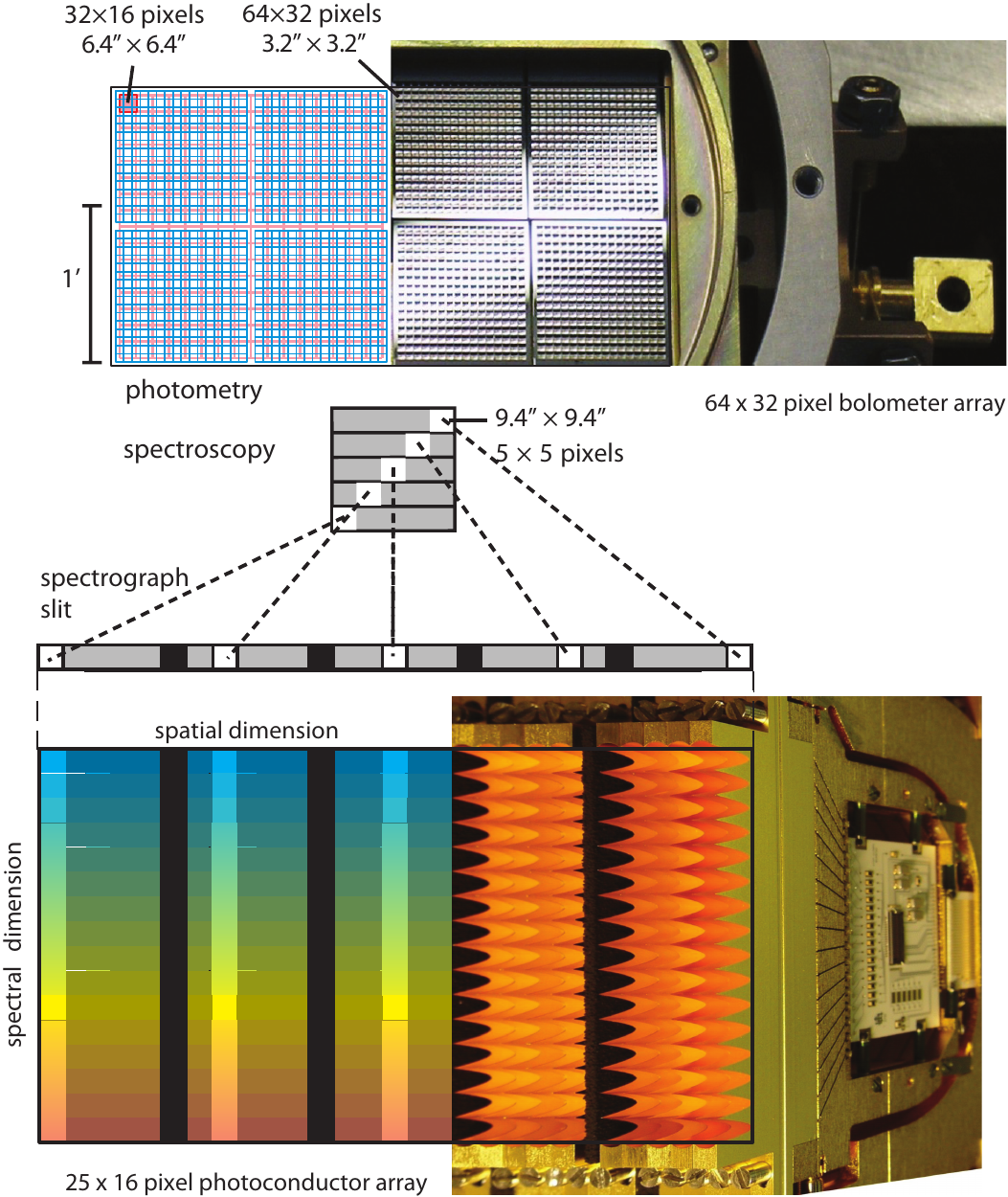}
   \caption
	{
\emph{Field splitter footprint and detectors}. 
A fixed mirror splits the focal plane into the photometry (Top) 
and spectroscopy (Bottom) channels of the instrument.  
The two photometer cameras have different magnification 
to cover the same field of view with a different pixel scale. 
In spectroscopy, an optical image slicer re-arranges the 
2-dimensional field along the entrance slit of the grating 
spectrograph such that for each spatial element in the field 
of view, a spectrum can be simultaneously observed with a 2D 
detector array. Right: The close-up photos show, at 
the appropriate scale, half of the blue/green bolometer array 
with its tiled monolithic sub-arrays (see \ref{sec:bol}) 
and part of the red photoconductor array with its area-filling
light-cones and CREs (see \ref{sec:photocond}). 
	}
     \label{fig:fov}%
\end{figure}

In an intermediate focus after the chopper, a fixed field mirror splits off the light for the
spectroscopy channel. The remaining part of the field of view
passes into the photometry channels. A ``footprint'' of the
focal-plane splitter is shown in Fig.~\ref{fig:fov}.

The calibrators are placed near the entrance to the
instrument, outside of the Lyot stop, to have approximately the same light path for observation and
internal calibration. The calibrator sources are grey-body sources providing FIR
radiation loads slightly above or below the telescope background,
respectively. They uniformly illuminate both, the field of view, and the Lyot stop, 
to mimic the illumination by the telescope.

The chopper  provides a maximum
throw of $6'$ on the sky; this
allows full separation of an ``object'' field from  a ``reference''
field. The chopper (\cite{krause06}) is capable of following
arbitrary waveforms with a resolution of $1''$ and delivers a
duty-cycle of $\sim$80\% at a chop frequency of 5~Hz.

	\subsection{Imaging photometer}

After the intermediate focus provided by the front optics, the
light is split into the long-wave (``red'') and short-wave (``blue'', ``green'')
channels by a dichroic beam-splitter with a transition wavelength
of $130\,\mu$m (design value) and is re-imaged with different magnification onto the
respective bolometer arrays.

The $32\times 16$ (red) or $64\times 32$ (blue/green) pixels in each array are used
to image the same field of view of $3.5'\times 1.75'$, the different magnification
providing full beam sampling at 90$\,\mu$m and 180\,$\mu$m,
respectively.  Projected onto the sky, pixel sizes are $6.4\arcsec \times 6.4\arcsec$ (red) and $3.2\arcsec \times 3.2\arcsec$ (blue/green), respectively. The red band (130--210\,$\mu$m) can be
combined with either the blue or the green
channel, 60--85$\,\mu$m or 85--130$\,\mu$m, for simultaneous observation.
Blue and green are selected by filter wheel. All
filters in PACS are implemented as multi-mesh filters and
provided by Cardiff University\footnote{P.A.R.\ Ade, Department of
Physics and Astronomy, University of Wales, Cardiff, UK} (\cite{ade06}).

	\subsubsection{Bolometer arrays}\label{sec:bol}

The PACS bolometers are filled arrays of square pixels which
allow instantaneous beam sampling.  $4 \times 2$ monolithic
sub-arrays of $16 \times 16$ pixels each are tiled together to form
the blue/green focal plane array (see Fig.~\ref{fig:fov}).
There remains a small gap between sub-arrays, 
which has to be filled by mapping methods (rastering, scanning).
Similarly, two sub-arrays of $16 \times 16$ pixels
are form the red focal plane array. 
The bolometer assemblies are kept at ``Level 0'' cryostat 
temperature ($\sim$1.65\,K).

The bolometer sub-arrays themselves
are mounted thermally isolated from
the surrounding 2 K structure and at an operating temperature of 0.3\,K.
This is provided by the closed cycle $^3$He sorption cooler (\cite{duband08}), designed to support uninterrupted operation of the bolometers for two days.  After this period, a
recycling is required. This process is usually carried out during
the daily telecommunication period of the satellite.
Bolometer cameras and cryo-cooler are mounted together in a self contained subunit
of the FPU. 

The cold readout 
electronics of the bolometers is split in two levels; a first stage is 
on the back of
the focal plane arrays, operating
at 0.3~K, and a second buffer stage runs at 2~K. The multiplexing readout samples 
each pixel at a rate of 40~Hz. 

The post-detection bandwidth (thermal/electrical) of the bolometers is $\sim$5~Hz at 
nominal bias and can be traded off against NEP;
the noise of the bolometer/readout system, however, has a strong 1/f component.  
To achieve optimum sensitivity, observations have to be executed such that the 
signal modulation due to chopping or scanning falls primarily in the frequency band 
from 1\,Hz to 5\,Hz, where close to background-limited performance is achieved.
Details on the bolometer design have been published in 
\cite{Agnese99}, 2003, \cite{Simoens04, Billot07}.

\subsection{Integral field spectrometer}

In the spectrometer, we use an integral field unit (IFU) 
feeding a Littrow-mount grating spectrometer to collect an
instantaneous 16-pixel spectrum with reasonable sampling and
baseline coverage for each of the 5 x 5 spatial image pixels.

The IFU concept has been selected because simultaneous
spectral and spatial multiplexing allows for the most efficient
detection of weak individual spectral lines with sufficient
baseline coverage and high tolerance to pointing errors without
compromising spatial resolution, as well as for spectral line
mapping of extended sources regardless of their intrinsic velocity
structure.

It is possible to operate both spectrometer detector arrays
simultaneously. For wide scans, full spectra can so be obtained in
both selected grating orders. In a spectral line mode in the
grating order-of-interest, the other array yields narrow-band
continuum data, or in suitably line-rich sources, serendipitous lines.

The spectrometer covers the wavelength
range 55--210$\,\mu$m, with simultaneous imaging
of a $47''\times 47''$ field of view in two grating orders, resolved into $5\times
5$ pixels. The spectral resolving power of 1000--4000 ($\Delta\rm{v} = 75$--300\,km/s) with an
instantaneous coverage of $\sim$1500\,km/s, depends on wavelength and order. A detailed
resolution curve is given in the \cite{pacs10} (2010). 

In the IFU, an image slicer employing reflective optics is used to
re-arrange the 2-dimensional field of view along a $1\times 25$
pixels entrance slit for a grating spectrometer, as schematically
shown in Fig.~\ref{fig:fov}.  A detailed description of the 
slicer optics, including a physical optics analysis, is given in a paper
on the similar SOFIA experiment FIFI LS (\cite{Looney03}).

The Littrow-mounted grating with a length of $\sim 300$\,mm is
operated in 1$^{\rm st}$, 2$^{\rm nd}$ or 3$^{\rm rd}$ order,
respectively, to cover the full wavelength range. Nominally, the 1$^{\rm st}$
order covers the range 105--210$\,\mu$m, the 2$^{\rm nd}$ order 72--105$\,\mu$m,
 and the 3$^{\rm rd}$ order 55--72$\,\mu$m (design values --
actual filter edges slightly deviate from these). Anamorphic collimating
optics expands the beam along the grating over a length required to
reach the desired spectral resolution. The grating is actuated by a
cryogenic motor (\cite{Renotte99}) which, together with arc-second precision
position readout and control, allows spectral scanning/stepping for
improved spectral flat-fielding and for coverage of extended wavelength
ranges. The settling time for small motions is $\leq 32$ms, enabling
wavelength-switching observations.

Anamorphic re-imaging optics after the grating spectrometer allows one
to independently match the dispersed image of the slit spatially to the 25 pixel columns and 
adjust the dispersion such that the square pixels of the detector
arrays well sample the spectral resolution.

The light from the 1$^{\rm st}$ diffraction order is separated from
the higher orders by a dichroic beam splitter and
passed on into two optical trains feeding the respective detector
array (stressed/unstressed) for the wavelength ranges 105--210$\,\mu$m and
55--105$\,\mu$m. The filter wheel in the
short-wavelength path selects  2$^{\rm nd}$ or 3$^{\rm rd}$ grating order.

\subsubsection{Photoconductor arrays}\label{sec:photocond}

The 25$\times$16 pixels Ge:Ga photoconductor arrays employed in the spectrometer
are a completely modular
design. 25 linear modules of 16 pixels each are stacked together to form a contiguous,
2-dimensional array. Each of the modules records an instantaneous spectrum of 16 pixels, as 
described above.

Light cones in front of the actual detector block
provide area-filling light collection in the focal plane. Details of the design of
both arrays are given in \cite{Kraft00,Kraft01,PogDet03}.

Responsivity measurements of both stressed and unstressed modules show sufficiently
homogeneous spectral and photometric response within each module and
between modules. Absolute responsivity calibration 
for optimum bias under in-orbit conditions is under way and will most likely give  
numbers of $\sim$10 A/W for the unstressed detectors and $\sim$40 A/W for
the stressed detectors. The detectors are operated at (stressed) or slightly above (unstressed) the 
``Level 0'' cryostat temperature ($\sim$1.65\,K).

Each linear module of 16 detectors is read out by a cryogenic amplifier
/multiplexer circuit (CRE) in CMOS technology (\cite{Merken04}). 
The readout electronics is integrated into the detector modules (see Fig.~\ref{fig:fov}), but operates at 
``Level 1'' temperature (3\ldots5\,K).

Measurements of the NEP of both arrays after integration into the instrument 
flight model at characteristic wavelengths and with representative flux 
levels have confirmed the performance 
measured at module level. Only a small fraction of pixels suffers from excess noise.  
Median NEP values 
are $8.9\times10^{-18}\rm{\,W/Hz^{1/2}}$ for the stressed and 
$2.1\times10^{-17}\rm{\,W/Hz^{1/2}}$ for the unstressed detectors, respectively.

The achievable in-orbit performance was expected to depend critically on
the effects of cosmic rays on the detector response. Proton irradiation
tests performed at the synchrotron source of the Universit\'e Catholique
de Louvain (Louvain la Neuve, Belgium) complemented by a $\gamma$-radiation test
programme at MPIA indicated that NEPs close to
those measured without irradiation should actually be achievable in
flight (\cite{Katterloher06,Stegmaier08}).

\subsection{Instrument control electronics and on-board data processing}

The warm electronics units of PACS on the satellite bus have several tasks: 
control the instrument, send the housekeeping and science data, and provide autonomy in the 
20 hour interval between the daily telemetry periods.
PACS is far too complex to be controlled by a single electronics unit. Therefore, the various 
functions have been grouped into different subunits.
	
The Digital Processing Unit (DPU) provides the interface of PACS to the
satellite and is therefore responsible for receiving and decoding
commands from the ground or the mission timeline on-board.
Decoded commands are forwarded by the DPU to the appropriate subsystems
for execution. The DPU monitors the correct execution of all commands
and raises errors should a failure occur. In the opposite data flow
direction, all housekeeping and science data from the various subsystems
of PACS are collected by the DPU, formatted into telemetry packets and
sent to the satellite mass memory.

The DEtector and MEchanism Controller (DECMEC) receives and handles all
low level commands to all PACS systems, except for the bolometers.
Also, tasks that need to be synchronous (e.g., detector readouts,
chopper motion, grating steps) are triggered from here. The DEC part
operates the photoconductor arrays and receives their raw data, which
arrive at 256 Hz for each pixel. It also receives the digitised
bolometer data from BOLC (see below). The data are then sent on to the
SPU (see below) for processing. The MEC part contains drivers for all
mechanisms and regulated temperatures in the FPU and generates most of
the instrument housekeeping data.

The BOLometer Controller (BOLC) operates the bolometer arrays and
provides a clock signal to MEC for the chopper synchronisation in
photometer mode. The digitised bolometer signals are sent to the SPU via
DECMEC for processing.

The Signal Processing Unit (SPU) reduces the raw data rate from the detector arrays, which
exceeds the average allowed telemetry rate of 130 kb/s by far. The SPU performs real-time reduction of
the raw data in the time domain, bit rounding, and subsequent 
lossless compression. The algorithms employed (\cite{Ottensamer08}) are optimised 
for each type of detector and observing mode to keep as true to the original raw data as possible.

\section{Observing modes}

Typical PACS Observation Days (OD) contain predominantly either photometer or
spectrometer observations to optimise the observing efficiency within
a photometer cooler cycle (see Sect.~\ref{performance}). After each
cooler recycling procedure (which takes about 2.5\,h), there are about
2.5\,ODs of PACS photometer -- prime or parallel mode -- observations possible.
Mixed days with both sub-instruments, e.g., to observe the same target in
photometry and spectroscopy close in time, are only scheduled in exceptional
cases.

\subsection{Photometer}

Three observing modes or Astronomical Observing Templates (AOT)
are validated on the PACS photometer side:
(i) Point-source photometry mode in chopping-nodding technique
(ii) Scan map technique (for point-sources, small and large fields)
(iii) Scan map technique within the PACS/SPIRE parallel mode.
The originally foreseen "small source mode" and "large raster mode"
in chopping-nodding technique are replaced by the scan-map technique
for better performance and sensitivity reasons.

All photometer configurations perform dual-band photometry with
the possibility to select either the blue (60--85\,$\mu$m) or the green
(85--125\,$\mu$m) filter for the short wavelength band, the red band 
(125--210\,$\mu$m) is always included. The two
bolometer arrays provide full spatial sampling in each band.

During an observation the bolometers are read-out with 40\,Hz,
but due to satellite data-rate limitations there are on-board
reduction and compression steps needed before the data is 
down-linked. In PACS prime modes the SPU averages 4
subsequent frames; in case of chopping the averaging process
is synchronised with the chopper movements to avoid averaging
over chopper transitions. In PACS/SPIRE parallel mode 8
consecutive frames are averaged in the blue/green 
bands and 4 in the red band. In addition to
the averaging process there is a supplementary compression
stage "bit rounding" for high gain observations required,
where the last $n$ bits of the signal values are rounded off. 
The default value for $n$ is 2 (quantisation step of 2$\cdot$10$^{-5}$\,V
or 4\,ADU) for all high gain PACS/SPIRE
parallel mode observations, 1 for all high gain PACS prime mode observations,
and 0 for all low gain observations.

The selection of the correct gain ("LOW" or "HIGH") is driven by source
flux estimates given by the observer. The switch to low gain is required
for the flux limits given in Table~\ref{photdynamic}.
Each observation -- chop-nod or scan-map -- can be repeated several times,
driven by the observer-specified repetition factor.

Each PACS photometer observation is preceded by a 30\,s chopped
calibration measurement executed during the target acquisition
phase\footnote{Note: In early mission phases (until OD\,150) long
photometer observations were interleaved with additional
calibration blocks}.
The chopper moves with a frequency of 0.625\,Hz between the
two PACS internal calibration sources. 19 chopper cycles are executed, each chopper
plateau lasts for 0.8\,s (32 readouts on-board) producing 8 frames in the down-link.
There are always 5\,s idle-time between the calibration block and the on-sky part
for stabilisation reasons.

\subsubsection{Chop-nod technique}

The PACS photometer chop-nod point-source mode uses the PACS chopper to move the source
by about 50\arcsec, corresponding to the size of about 1 blue/green
bolometer matrix or the size of about half a red matrix, with a chopper frequency of 1.25\,Hz.
The nodding is performed by a satellite movement of the same amplitude but
perpendicular to the chopping direction.

On each nod-position the chopper executes 3$\times$25 chopper cycles.
The 3 sets of chopper patterns are either
on the same array positions (no dithering) or on 3 different array positions
(dither option). In the dither-option the chopper pattern is displaced in $\pm$Y-direction
(along the chopper direction) by about 8.5\arcsec (2 2/3 blue pixels or
1 1/3 red pixels). Each chopper plateau lasts for 0.4\,s (16 readouts on-board) producing
4 frames per plateau in the down-link. The full 3$\times$25 chopper cycles per nod-position
are completed in less than 1 minute. The pattern is repeated on the second nod-position.
In case of repetition factors larger than 1, the nod-cycles are repeated in the following
way (example for 4 repetitions): nodA-nodB-nodB-nodA-nodA-nodB-nodB-nodA to minimise
satellite slew times.

The achieved sensitivities (see Table~\ref{photcal}) are worse by a factor 1.5-2 compared to
the preflight prediction, due to different operating parameters.

Despite the degraded sensitivity this mode has advantages for intermediately bright
sources in the range 50\,mJy to about 50\,Jy: a small relative pointing error
(RPE) of 0.3\arcsec and high photometric reliability.

\subsubsection{Scan technique}

\begin{figure}
   \centering
   \includegraphics[width=0.87\columnwidth]{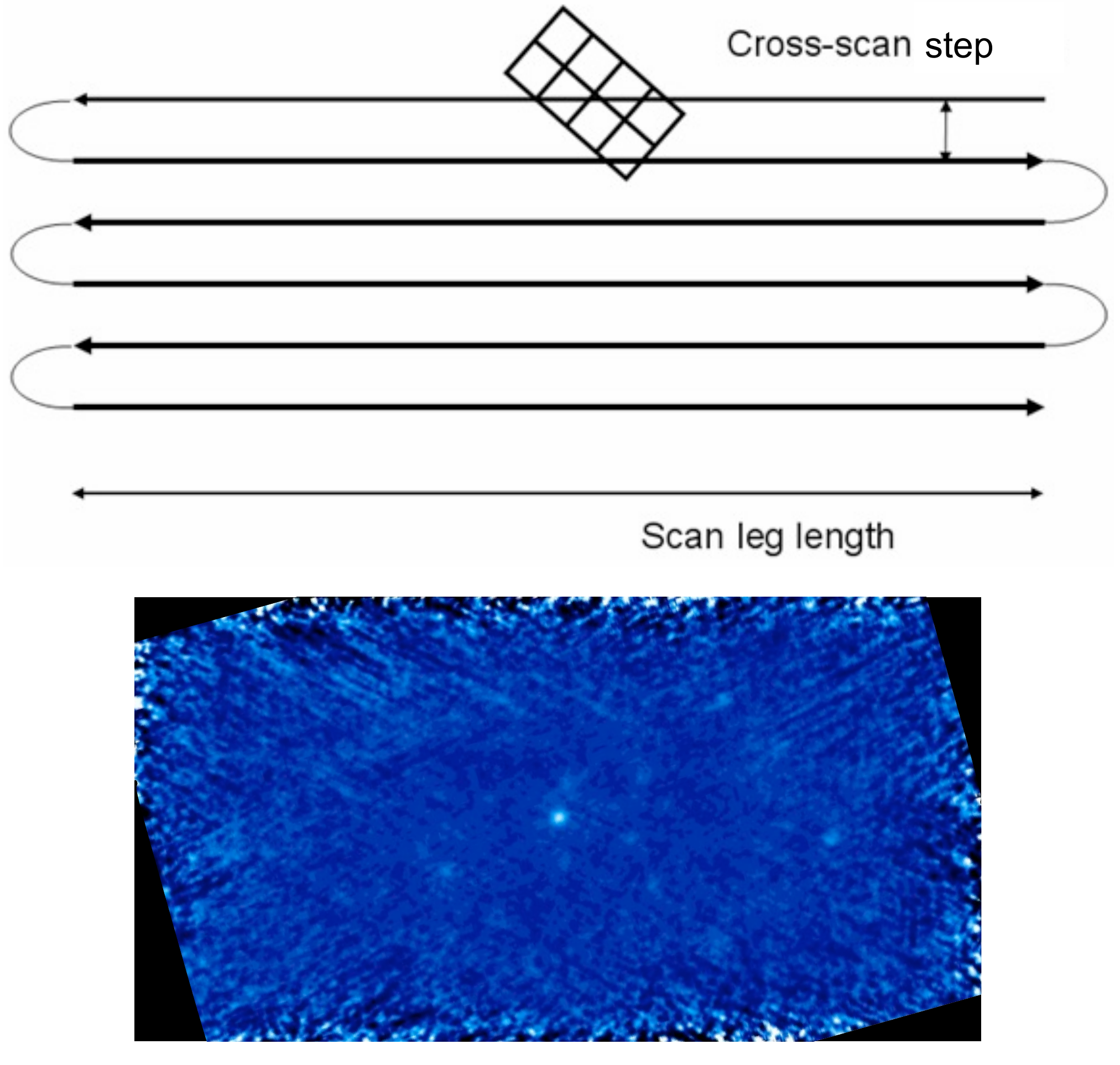}
      \caption{Top: Example of PACS photometer scan map. Schematic of a scan map with 6 scan
           line legs. After the first line, the satellite turns left and continue with
           the next scan line in the opposite direction. The reference scan
           direction is the direction of the first leg.
           Bottom: combined cross-scan mini-maps for HD\,159330 ($\sim$30\,mJy at 100\,$\mu$m.)
              }
         \label{scanmap1}
   \end{figure}

The scan-technique is the most frequently used {\em Herschel} observing mode.
Scan maps are the default to map large areas of the sky, for galactic as well
as extragalactic surveys, but meanwhile they are also recommended for small
fields and even for point-sources. Scan maps are performed by slewing the
spacecraft at a constant speed along parallel lines (see Fig.~\ref{scanmap1}).
Available satellite speeds are 10, 20, 60\arcsec/s in PACS prime mode
and 20, 60\arcsec/s (slow, fast) in PACS/SPIRE parallel mode. The number
of satellite scans, the scan leg length, the scan leg separation, and the orientation
angles (in array and sky reference frames) are freely selectable by the observer.
Via a repetition parameter the specified map can be repeated $n$ times. The performance
for a given map configuration and repetition factor can be evaluated beforehand 
via sensitivity estimates and coverage maps in HSPOT\footnote{The \emph{Herschel}
observation planning
tool is available from {\tt http://herschel.esac.esa.int/}}. The PACS/SPIRE parallel mode
sky coverage maps are driven by the fixed 21\arcmin \ separation between the
PACS and SPIRE footprints. This mode is very inefficient for small fields, the shortest
possible observation requires about 45\,min observing time.

During the full scan-map duration the bolometers are constantly read-out with
40\,Hz, allowing for a complete time-line analysis for each pixel in the
data-reduction on ground. A combination of two different scan directions is recommended
for a better field and PSF reconstruction.

Most of the PACS prime observations are performed with a 20\arcsec/s scan speed
where the bolometer performance is best and the pre-flight sensitivity estimates are met. 
For larger fields observed in instrument reference frame there is an
option to use "homogeneous coverage" which computes the cross-scan
distance in order to distribute homogeneously the time spent on
each sky pixel in the map.

For short scan legs below about 10\arcmin \ the efficiency of this mode
drops below 50\% due to the relatively long time required for the
satellite turn-around (deceleration, idle-time, acceleration)
between individual scan legs, which takes about 20\,s for small leg separations
of a few arcseconds.
Nevertheless, this mode has an excellent performance for very small fields
and even for point-sources (see Sect.~\ref{performance} and Table~\ref{photcal}). 

The advantages of the scan mode for small fields are the better
characterisation of the source vicinity and larger scale structures in the 
background, the more homogeneous coverage inside the final map, the higher
redundancy with respect to the impact of noisy and dead pixels and the
better point-source sensitivity as compared to a chop-nod observation
of similar length.

\subsection{Spectrometer}

There are three validated PACS spectrometer observing modes: {\em Chopped line 
spectroscopy} for single lines on sources with a clean background 
within 6\arcmin, {\em chopped range spectroscopy} for spectra over larger
wavelength ranges on sources with a clean background within 6\arcmin, 
and the {\em wavelength switching mode}, for single lines on extended
sources without clean background for chopping. An {\em unchopped observing mode} 
for larger wavelength ranges on extended sources is planned. 

The three observing modes can be used in a single pointing, or repeated in 
a raster pattern on the sky.  There are two sets of recommended raster patterns
for mapping with full beam sampling: one for compact sources, which fit within the
instantaneous FOV of the spectrometer, and one for more extended sources. 
For compact sources, the recommended pattern is a $3\times3$ raster with
a 3\arcsec \ step size in the blue bands and a $2\times2$ raster with a 4.5\arcsec \ 
step size in the red band. For extended sources, $m\times n$ rasters
with approximately 5/3 pixel step sizes in the blue bands and
approximately 5/2 pixel step sizes in the red bands are recommended (see
also \cite{pacs10} 2010)

All spectrometer observing modes include a 
calibration block, a modulated measurement of
the two internal calibration sources with the grating in a fixed position.
The two sources are heated to different temperatures, 
hence provide different signal levels.  
The grating position is chosen to measure a reference wavelength in the 
bands that are measured in the sky observation.  
Table~\ref{table:keyWavelengths} lists these calibration block wavelengths.

The calibration block measurement starts during the slew of the spacecraft
to the target in order to optimise the use of observing time. 

\begin{table}
\begin{minipage}[t]{\columnwidth}
\caption{Spectrometer calibration block key wavelengths}
\label{table:keyWavelengths}
{\centering 
\begin{tabular}{c c c} 
\hline\hline
Bands     &  Blue wavelength & Red wavelength \\
\hline
B2A / R1  &  60              & 120 \\
B3A / R1  &  60              & 180 \\
B2B / R1  &  75              & 150 \\
\hline
\end{tabular}}
\vspace{2mm}
\end{minipage}
The wavelengths
observed in the spectrometer calibration block depend
on the spectral bands visited in the rest of the observation.
\vspace{-2mm}
\end{table}

Based on the continuum and line flux estimates entered by the observer,
the expected maximum photoconductor signal level is estimated by the
observing logic.  
For range spectroscopy, the expected flux at the maximum response is 
extrapolated via a Rayleigh-Jeans law from the reference wavelength 
and corresponding flux estimate.  
The appropriate integrating capacitance of the CRE is then chosen for the 
entire observation to avoid saturation.

\subsubsection{Chopped line scan spectroscopy}

The chopped line spectroscopy mode can contain up to 10 spectral 
line scans across different bands with the same fixed order selection 
filter wheel position. The diffraction grating is rotated over 43 to 48 steps
so that every spectral pixel homogeneously samples one resolution element 
at least 3 times.  This scan is repeated in two directions.  This up/down scan
can be repeated up to 10 times for a single line.
Table~\ref{table:lineScanSpectralCoverage} shows the wavelength range 
covered in the different bands. 

At every grating position, the detector signal is modulated between on and off
source via an on-off-off-on-on-off-off-on chopping pattern.  At every 
chopper plateau two 1/8 second integrations of the photoconductor signals 
are recorded. The observer can choose a chopper throw of 6\arcmin, 3\arcmin or 1\arcmin.  

The sequence of line scans is repeated at two nod positions of the
telescope.  In the second nod position, the source is located on the 
off chopping position of the first nod in order to be able to determine 
the difference in telescope background at the two chop positions.
The nod sequence can be repeated within one observation to increase
the depth of the observation.

\begin{table}
\begin{minipage}[t]{\columnwidth}
\caption{The wavelength range seen in a nominal line scan}
\label{table:lineScanSpectralCoverage}
{\centering                 
\begin{tabular}{c c c c c c c} 
\hline\hline
Band  & $\lambda$ & \multicolumn{2}{c}{Full range} & Every pixel &  \multicolumn{2}{c}{FWHM} \\
      & ($\mu$m)   & (km/s)     & ($\mu$m)   & ($\mu$m)         & ($\mu$m)  &(km/s) \\
B3A   & 55         & 1880       & 0.345      & 0.095            & 0.021      & 115 \\
B3A   & 72         & 800        & 0.192      & 0.053            & 0.013       & 55 \\
B2B   & 72         & 2660       & 0.638      & 0.221            & 0.039      & 165 \\ 
B2B   & 105        & 1040       & 0.364      & 0.126            & 0.028     & 80 \\
R1    & 105        & 5210       & 1.825      & 0.875            & 0.111      & 315 \\
R1    & 158        & 2870       & 1.511      & 0.724            & 0.126      & 240 \\
R1    & 175        & 2340       & 1.363      & 0.654            & 0.124      & 210 \\
R1    & 210        & 1310       & 0.92       & 0.441            & 0.098       & 140 \\
\hline
\end{tabular}}
\vspace{2mm}
\end{minipage}
This range varies over  the spectral bands. The column `every pixel' refers
to the range that is seen by every spectral pixel.
\vspace{-2mm}
\end{table}

The line spectroscopy AOT provides a bright line mode. The sampling
step size is identical to the nominal, faint line mode, but the 
number of steps in the scan is limited to 16.  Therefore, in the bright
line mode the range scanned is 1/3 of the ranges listed in 
Table~\ref{table:lineScanSpectralCoverage}.

\subsubsection{Chopped range scan and SED spectroscopy}

The chopped range scan AOT allows one to observe one or several spectral 
ranges.  The mode provides two spectral sampling depth.  The deep 
sampling uses the same sampling step size as the line spectroscopy 
mode and is typically used to measure broadened lines or ranges with
several lines.   The coarser Nyquist sampling depth provides a sampling 
of at least one sample in every half resolution element by one of the 
spectral pixels.  Chopping frequency and scheme as well as the structure of the observation
are the same as in line spectroscopy: a sequence of range scans is repeated
in two nod positions, switching the source position between the two
chopper positions.

Pre-defined ranges are foreseen to measure full PACS SEDs. These are
measured in Nyquist sampling depth.

Repetitions of the same range within the nod position in Nyquist sampling
depth are offset in wavelength, so that the same wavelength is sampled by 
different spectral pixels in every repetition.

\subsubsection{Wavelength switching spectroscopy}

The wavelength switching technique/mode is an alternative to the 
chopping/nodding mode, if by chopping to a maximum of 6\arcmin, 
the OFF position field-of-view cannot be on an
emission free area, for instance in crowded areas.

In wavelength switching mode, the line is scanned with the same
grating step as in chopped line spectroscopy, i.e., every spectral pixel
samples at least every 1/3 of a resolution element.  In wavelength
switching we refer to this step as a dither step.  At every dither  
step, the signal is modulated by moving the line over about half of the 
FWHM.  This allows one to measure a differential line profile, canceling
out the background.
The modulation on every scan step follows an AABBBBAA 
pattern, where A is a 
detector integration at the initial wavelength, and B is a detector integration
at the wavelength switching wavelength.  This cycle is repeated 20 times in 
one direction, and repeated in the reverse wavelength direction.
The switching amplitude is fixed for every spectral band.

In order to reconstruct the full power spectrum, a clean off-position is 
visited at the beginning and the end of the observation.  On this position
the same scan is performed.  In between, the scan is performed at two or more
raster positions.

\section{Calibration}

Prior to launch, the instrument has been exposed to an extensive ground
based test and calibration programme. The resulting instrument
characterisation has been determined in a specific test cryostat
providing all necessary thermal and mechanical interfaces to the
instrument. In addition to the PACS FPU, the cryostat hosted a telescope
simulation optics, cryogenic far-infrared black-body sources, and a set
of flip and chop mirrors. External to the test cryostat a methanol laser
setup (\cite{inguscio86}), movable hole masks with various diameters,
illuminated by an external hot black-body and a water vapour absorption
cell, have been available for measurements through cryostat windows. The
results of this instrument level test campaign are summarised in
\cite{alpog08} and served as a basis for the in-flight calibration in the
performance verification phase of satellite and instruments.

\subsection{Photometer}

\begin{figure}
   \centering
   \includegraphics[width=0.98\columnwidth]{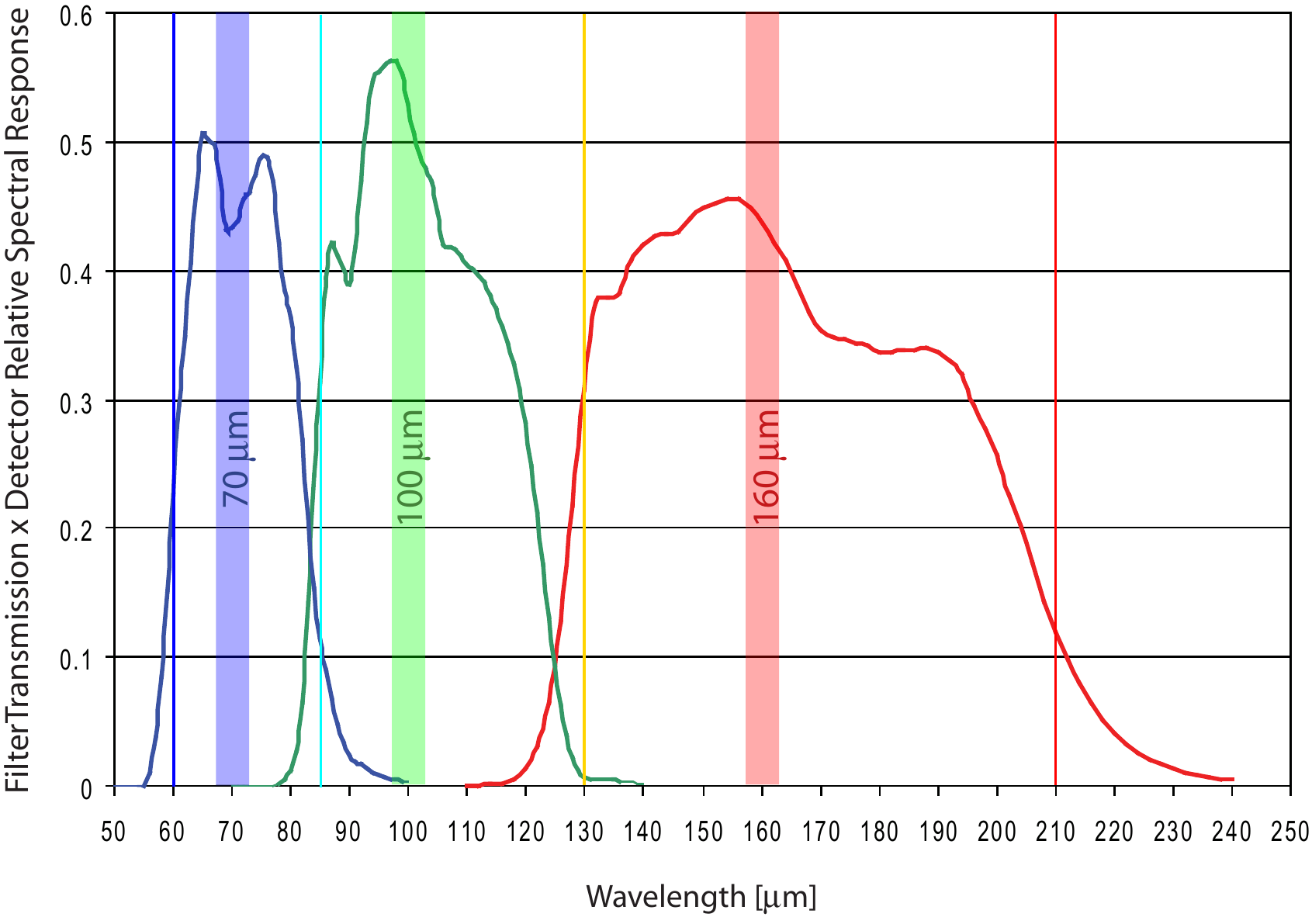}
      \caption{Effective spectral response of the filter/detector chain of the PACS photometer in its three bands.
              }
         \label{photbands}
   \end{figure}

The absolute flux calibration of the photometer is based on models of fiducial stars
($\alpha$\,Cet, $\alpha$\,Tau, $\alpha$\,CMa, $\alpha$\,Boo,
 $\gamma$\,Dra, $\beta$\,Peg; \cite{Dehaes10}) and thermophysical
models for a set of more than 10 asteroids (\cite{MulLag98},
2002), building up on similar approaches for ISOPHOT (\cite{Schulz02}) and
Akari-FIS (\cite{Shira09}).   Together they cover a flux range from below 100\,mJy up to 300\,Jy.
Both types of sources agree very well in all 3 PACS bands, and the
established absolute flux calibration is consistent within 5\%. 
The quoted photometric accuracies in Table~\ref{photcal} can thus be considered conservative.
Neptune and Uranus with flux levels of several hundred Jansky are
already close to the saturation limits, but have been used
for flux validation purposes.  At those flux levels, a reduction in response of up to 10\% has been observed.  For comparison, the latest FIR flux model of Neptune is considered to be accurate to better than 5\% (\cite{Fletch10}).
No indications of any near- or mid-infrared filter leakage could be identified. 
Required colour corrections for
the photometric PACS reference wavelengths (70, 100, 160\,$\mu$m) which
have been determined from the photometer filter transmission curves and
bolometer responses (see Fig.~\ref{photbands}) are quite small.
Suitable correction factors for a wide sample of SED
shapes can be read from Table~\ref{colorcor} but are also available from
within the \emph{Herschel} Interactive Processing Environment.
Absolute flux calibration uncertainties should improve over
the mission, with better statistics of available celestial calibration
observations.

\begin{table}
\begin{minipage}[t]{\columnwidth}
\caption{Photometric colour corrections} 
\label{colorcor}      
\centering                          
\renewcommand{\footnoterule}{} 
\begin{center}
\begin{tabular}{lcccccc}
 \hline \noalign{\smallskip}
BB temp. &  CC\_70 &  CC\_100 & CC\_160 \\
\noalign{\smallskip} \hline \noalign{\smallskip}
BB (10000 K) &  1.02   &  1.03    & 1.07  \\
BB (5000 K)  &  1.02   &  1.03    & 1.07  \\
BB (1000 K)  &  1.01   &  1.03    & 1.07  \\
BB (500 K)   &  1.01   &  1.03    & 1.07  \\
BB (250 K)   &  1.01   &  1.02    & 1.06  \\
BB (100 K)   &  0.99   &  1.01    & 1.04  \\
BB (50 K)    &  0.98   &  0.99    & 1.01  \\
BB (20 K)\footnote{ Colour corrections for sources with temperatures
                 below 20\,K can become quite significant, in particular at 70\,$\mu$m.
                
                Photometric reference spectrum:
                 $\nu F_{\nu}=\lambda F_{\lambda}=\rm{const.}$.
                 PACS bolometer reference wavelengths:
                 {70.0, 100.0, 160.0\,$\mu$m}. In order to obtain a monochromatic flux density
                  one has to divide the measured
                 and calibrated band flux by the above tabulated values.   
}                    &  1.22   &  1.04    & 0.96  \\
BB (15 K)    &  1.61   &  1.16    & 0.99  \\
BB (10 K)    &  3.65   &  1.71    & 1.18  \\
\noalign{\smallskip} \hline \noalign{\smallskip}
\noalign{\smallskip} \hline \noalign{\smallskip}
Power law ($\nu^{\beta}$)  &  CC\_70 &  CC\_100 & CC\_160 \\
\noalign{\smallskip} \hline \noalign{\smallskip}
$\beta = -3.0 $  &  1.04  & 1.04  & 1.06 \\
$\beta = -2.0 $  &  1.02  & 1.01  & 1.02 \\
$\beta = -1.0 $  &  1.00  & 1.00  & 1.00 \\
$\beta =  0.0 $  &  1.00  & 1.00  & 1.00 \\
$\beta =  1.0 $  &  1.00  & 1.01  & 1.03 \\
$\beta =  2.0 $  &  1.02  & 1.03  & 1.08 \\
$\beta =  3.0 $  &  1.04  & 1.07  & 1.14 \\
\end{tabular}
\end{center}
\end{minipage}
\end{table}

   \begin{figure}[h]
   \centering
   \includegraphics[width=0.9\columnwidth]{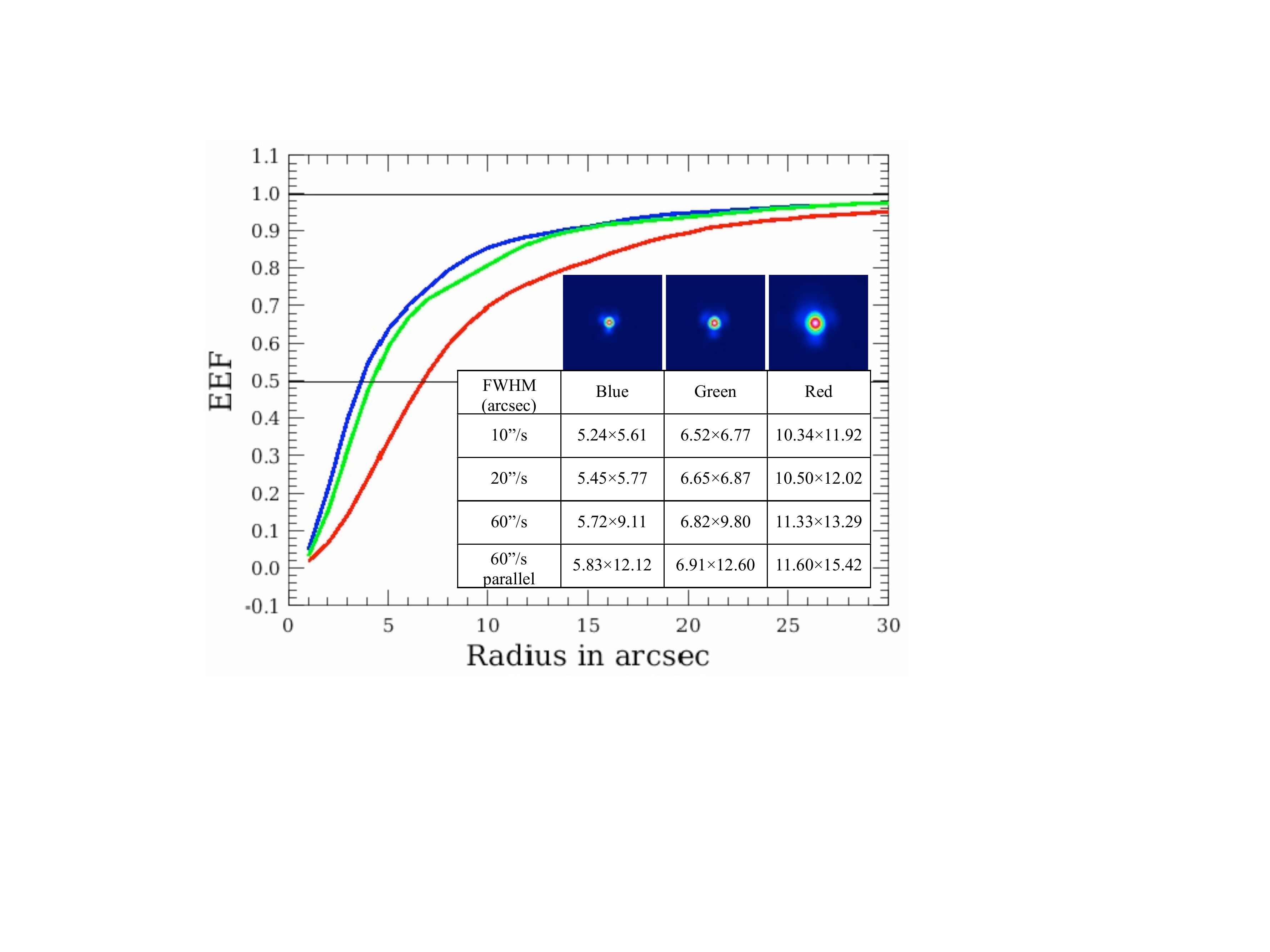}
   \vspace{1 mm}
      \caption{Photometer encircled energy function for blue, green and red photometer bands,
      for 10\arcsec /s scan speed.
       The inset tabulates observed beam sizes (FWHM) as a function of scan speed.  The bolometer time
       constant leads to an elongation in scan direction.  Note that this
        elongation is additionally increased in SPIRE/PACS parallel observations, due
        to data compression (see also Sect.~\ref{phot_performance}).  The beam maps, shown for illustration, were taken at a 10\arcsec /s scan speed.}
         \label{photeef}
   \end{figure}

The photometer focal plane geometry, initially established on ground by
scanning a back-illuminated hole mask across the bolometer arrays, has
been adapted to the actual telescope by optical modeling. The in-flight
verification required a small change in scale and a slight rotation to
fit to the results of a 32$\times$32 raster on $\alpha$ Her. In
particular these measurements also show that the possibilities to
further improve the calibration of detailed distortions are limited by
short-term pointing drifts of the satellite. Residual measured
inter-band offsets between blue/green (0.3\arcsec) and green/red (1.2\arcsec) have
been characterised and are implemented in the astrometric processing
chain. Current best point spread functions (PSF) have been determined on
asteroid Vesta. The observed 3-lobe structure of the PSF (Fig.~\ref{photeef} and \cite{pilbratt10}) can be
explained qualitatively by the secondary mirror support structure and
has been verified in detail by ray tracing calculations taking the
telescope design and known wave-front errors into account.
The spatial resolution, expressed as encircled
energy as a function of angular separation from PSF
centre (Fig.~\ref{photeef}), is in reasonable agreement with expectations from telescope and
instrument design.

\subsection{Spectrometer}

The wavelength calibration of the PACS spectrometer relates the grating angle to the
central wavelength ``seen'' by each pixel.
Due to the finite width of the spectrometer slit, a
characterisation of the wavelength scale as a function of (point) source
position within the slit is required as well. 

The calibration derived
from the laboratory water vapour absorption cell is still valid
in-flight. For ideal extended sources the required accuracy of better
than 20\% of a spectral resolution is met throughout all bands.
While at band borders, due to leakage effects and lower S/N,
the RMS calibration accuracy is closer to 20\%,
values even better than 10\% are obtained in band centres.
However, for point sources the wavelength calibration may be dominated by
pointing accuracy.

   \begin{figure}[b]
   \centering
   \includegraphics[width=0.8\columnwidth]{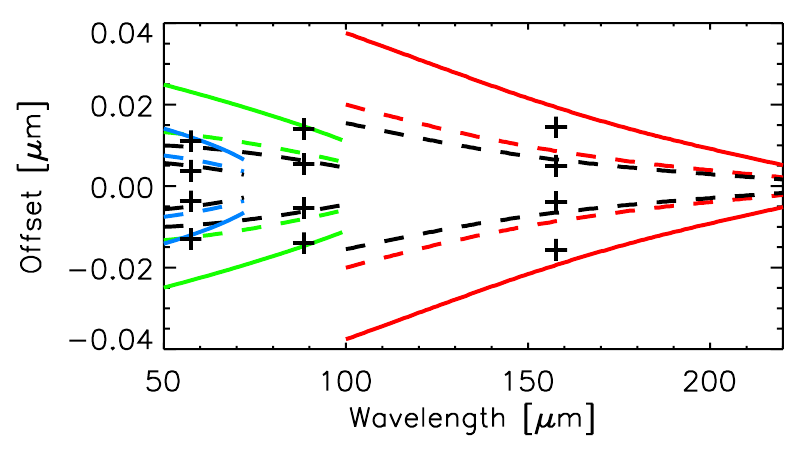}
	\caption{Calculated wavelength offsets for point source positions: at
	the slit border (solid colour lines), for typical pointing errors up to
	2\arcsec (dashed colour lines) and measured line centre offsets for $\pm$1.5\arcsec
	(dashed black line and crosses) and slit border (black crosses) for
	three spectral lines on the point like planetary nebula IC2501. }
         \label{SpecWaveCal}
   \end{figure}

Three 4$\times$4 raster observations (3\arcsec step size in instrument
coordinates) of the point-like planetary nebula IC2501 on the atomic
fine structure lines [N\,III] (57\,$\mu$m), [O\,III] (88\,$\mu$m) and
[C\,II] (158\,$\mu$m) have been compared with predictions from instrument
design. Figure~\ref{SpecWaveCal} shows good agreement between measured
spectral line centre positions and the predicted offsets from the
relative source position in the slit. Observed wavelength offsets for
individual point source observations are therefore expected to fall
within the dashed colour lines, given the nominal pointing uncertainty of
the spacecraft (\cite{pilbratt10}).

\begin{figure}
   \centering
   \includegraphics[width=0.74\columnwidth]{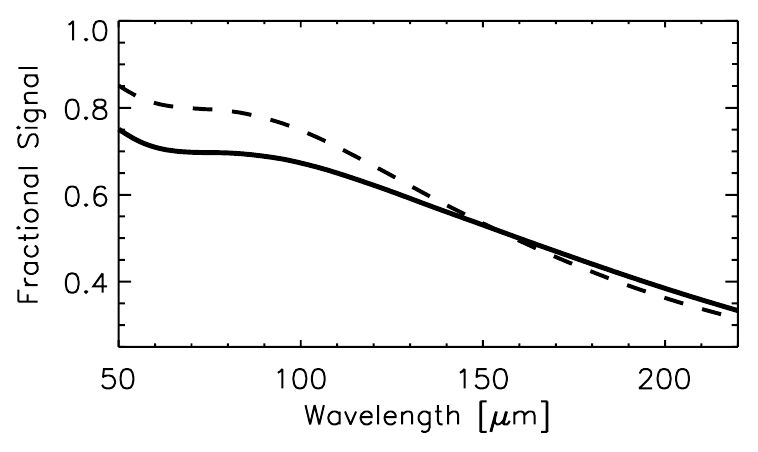}
	\caption{Point source correction: The fraction of signal seen by the
	central PACS spectrometer spaxel. Dashed line: theoretical calculation
	from idealised PSF; solid line: 3$^{\rm rd}$ order polynomial fit to results from
	rasters on Neptune.
              }
         \label{fluxcal}
   \end{figure}

The pre-launch flux calibration of the spectrometer is based on
measurements with the laboratory test
cryostat and its test optics. Significant changes to this calibration in
flight originate from a major response change of the Ge:Ga
photoconductors due to the radiation environment in the
L2 orbit, a retuning of the detector bias for optimum sensitivity, and
the telescope efficiency. On shorter time scales (one to few hours) the
Ge:Ga detectors also show slight drifts in absolute response. Each
observation is therefore preceded by a short exposure on the internal
calibration sources, which will allow to derive a detector pixel
specific relative correction factor. While the laboratory blackbodies
could be considered as ideal extended sources covering the entire
47\arcsec$\times$47\arcsec \,field of view of the spectrometer, celestial calibration
standards are typically point sources. Observations of point-like
targets in lower flux regimes can basically sample only the central
spatial pixel. A simple diffraction calculation (circular aperture with
central obscuration) has been initially adopted to correct point-source
fluxes. An improved point source correction has then been derived as a
result of extended 100\arcsec$\times$100\arcsec \,rasters on Neptune. The results are
presented in Fig.~\ref{fluxcal}. After correcting for the increased
absolute response (factors 1.3 and 1.1 for blue and red spectrometer
respectively), taking the point source correction into account, but
without drift correction yet, the absolute flux calibration uncertainty
is of order 30\%. Improved statistics on results from
celestial standards and implementation and use of the results of the
internal calibration block in the processing pipeline are in progress.

\begin{figure}
   \centering
   \includegraphics[width=0.6\columnwidth]{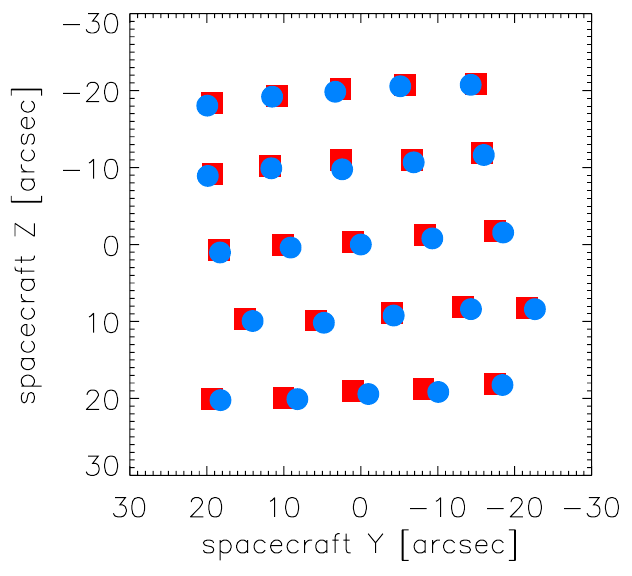}
      \caption{Spectrometer field of view for blue (circles) and red (squares) 
       spaxels in spacecraft Y and Z coordinates for chopper position zero.
              }
         \label{SpecFOV}
   \end{figure}

The spatial calibration of the PACS spectrometer section consists of the
detailed characterization of the relative locations on the sky of the
5$\times$5 spatial pixels (``spaxels'') in the blue and red sections and for all
operational chopper positions ($\pm$3', $\pm$1.5', $\pm$0.5', 0').
Detailed extended rasters on point sources (HIP21479 and Neptune) have
been carried out during PV phase at a few wavelengths and the resulting
spaxel geometries are stored as calibration files within the processing
software. Figure~\ref{SpecFOV} shows, as an example, the result for chopper
position zero in relative spacecraft units with respect to the virtual
aperture of the PACS spectrometer, which is defined as the central pixel
of the blue field of view. Asymmetrical optical distortions between
chopper on and off positions cause unavoidable slight misalignment
($\leq$2\arcsec) for individual spaxels between spacecraft nod A and B
within the double differential data acquisition scheme.

\begin{figure}
   \centering
   \includegraphics[width=0.9\columnwidth]{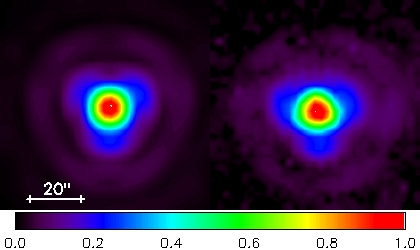}
	\caption{Calculated spectrometer PSF at 124\,$\mu$m (left) and
	measurement on Neptune (right) done at the same wavelength. Both are normalised to
	the peak and scaled by square-root, to enhance the faint wing pattern. The calculation
	includes the predicted telescope wave-front error, which dominates the overall
	aberrations.
              }
         \label{SpecPSF}
   \end{figure}

A further result of extended rasters on Neptune has been the
verification of the point spread function of the spectrometer.
Remarkable agreement with predictions from telescope and instrument
modeling has been found. A measurement for a typical spatial pixel of
the PACS spectrometer can be compared in Fig.~\ref{SpecPSF} to a
convolution of a calculated PSF (from actual telescope model including known
wave-front errors) with a square pixel of 9.4\arcsec$\times$9.4\arcsec.

%

%

\section{In-orbit performance}
\label{performance}

The instrument tests executed during
\emph{Herschel}'s commissioning phase successfully verified all critical
thermal and mechanical aspects. The PACS mechanisms and calibration
sources have been tuned to zero gravity conditions and are fully
functional. After passing through several weeks of post-launch
stabilisation, all thermal interface temperatures reached equilibrium
values close to or even better than expectations.

Following the satellite commissioning phase a comprehensive
characterisation and calibration programme of the instruments has been
executed within the performance verification phase of the \emph{Herschel}
mission. As a key pre-requisite to this programme, all PACS detector
supply voltages and heater settings had to be optimised for the thermal
and space radiation environment as encountered at L2 as well as for the
actual far-infrared thermal background emission caused by the 
temperature and emissivity of the telescope. The measured glitch rates
on the different PACS detectors were compatible with expectations at
solar minimum (Ge:Ga photoconductors: 0.08-0.2\,hits/s/pixel;
Bolometers: $\sim$1\,hit/min/pixel). The observed far-infrared telescope
background, mainly determined by M1 and M2 temperatures and the
telescope emissivity (see~\cite{fischer04}) was very close to
predictions during the first PACS spectrometer measurements.

The measured in-flight performance of the $^3$He cooler for the bolometers
substantially exceeds the required hold time of two days. The
cooler hold time, $t_{hold}$, as a function of $t_{on}$, the 
time during which the bolometers are actually powered-on is: 
~\mbox{$t_{hold} = 72.97{\rm \,h} - 0.20\times t_{on}$}.

Since the blue PACS photometer offers the highest spatial resolution
imaging on-board \emph{Herschel}, it has been used for determining the
spacecraft instrument alignment matrix (SIAM) defining the absolute
location of all the instruments' virtual apertures and the \emph{Herschel} absolute (APE) and
relative (RPE) pointing errors (\cite{pilbratt10}).  

\subsection{Photometer}
\label{phot_performance}

The performance of the photometer expressed as point source sensitivity
is listed in Table~\ref{photcal}. These numbers are derived from actual science observations.  
Our best estimate of the confusion-noise free sensitivity for deep integration comes from the scan map, where -- at the quoted levels -- source confusion is negligible in the blue and green bands, and still minor in the red band.
The sensitivity numbers in the other cases -- corrected for different integration times -- are fully consistent, with the exception of the red band in the parallel mode observation, which is from a shallow galactic plane survey, where source confusion already dominates the noise.

The upper flux limits of
the photometer, with the readout electronics in its (default) high-gain setting, are given in Table~\ref{photdynamic}.
Above these flux levels the gain will be set to "low",  which gives an extra headroom of $\sim$20\%.

\begin{table}
\begin{minipage}[t]{\columnwidth}
\caption{Bolometer readout saturation levels (high-gain setting)}       
\label{photdynamic}      
\centering                          
\renewcommand{\footnoterule}{} 
\begin{tabular}{c c c}        
\hline\hline                 
Filter &  Point source [Jy] & Extended source [GJy/sr]\\ 
\hline                        
   Blue    & 220  &  290     \\      
   Green & 510  &  350    \\
   Red    & 1125  &  300    \\ 
\hline                                   
\end{tabular}
\end{minipage}
\end{table}
\begin{table}
\begin{minipage}[t]{\columnwidth}
\caption{Absolute photometric uncertainty and point source sensitivity}       
\label{photcal}      
\centering                          
\renewcommand{\footnoterule}{} 
\begin{tabular}{c c c c c }        
\hline\hline                 
Band & Uncertainty\footnote{derived from 33, 23 and 51 targets for blue,
green and red point source observations respectively} & Point
source\footnote{The noise values quoted in the table are instrument limited for the blue
and green filter. The value for the red filter already contains
confusion noise contributions (see text).} & Scan map$^b$ & Parallel$^b$\\
         &                    &    mode                               & [$10'\times15'$] & [120'$\times$120']\\
        &                     & $5\sigma$/1h         &  $5\sigma$/30h &$1\sigma$/3h\\ 
\hline                        
   Blue    & $\pm$10\%  &  4.4\,mJy    & 3.7\,mJy   & 19.8\,mJy\\      
   Green & $\pm$10\%  &  5.1\,mJy    & 5.0\,mJy   & n.a.\\
   Red    & $\pm$20\%   &  9.8\,mJy & 9.5\,mJy   & 116\,mJy\\ 
\hline                                   
\vspace{-5mm} 
\end{tabular}
\end{minipage}
\end{table}


SPIRE/PACS parallel mode observations, typically in use on very large
fields require additional reduction (signal averaging to 200ms instead
of 100ms time bins) of blue photometer data. Together with scan speeds
of 60\arcsec/s, significant elongations of PSFs in such maps have been measured (Fig.~\ref{photeef}),
as expected.

The presence of faint ($\leq$1\%) optical ghosts and electrical crosstalk was
known from instrument level tests and has been reproduced in-flight.
Reflections from bright sources by the telescope structure are also known to produce
stray-light in the PACS field of view (\cite{pilbratt10}).

Accurate synchronisation between PACS detector data and
spacecraft pointing data is crucial for aligning forward and backward
directions of scan map observations and for superposing scan maps
obtained in orthogonal directions. There seems to be a systematic delay
of 50ms, by which the satellite pointing information is lagging the PACS data.
Compensation of this effect is not part of the pipeline and needs to be done ``manually''.

\subsection{Spectrometer}

The spectral resolution of the instrument, measured in the laboratory
with a methanol far-infrared laser setup (\cite{inguscio86}), follows
closely the predicted values of the 
\cite{pacs10}. Spectral lines in celestial standards (planetary
nebulae, HII regions, planets, etc.) are typically doppler broadened to
10-40 km/s, which has to be taken into account for any in-flight
analysis of the spectral resolution. The double differential chop-nod
observing strategy leads to slight wavelength shifts of spectral profile
centres due to the finite pointing performances. Averaging nod A and nod
B data can therefore lead to additional profile broadening causing
typical observed FWHM values that are up to $\sim$10\% larger.
     
\begin{figure}
   \centering
   \includegraphics[width=0.86\columnwidth]{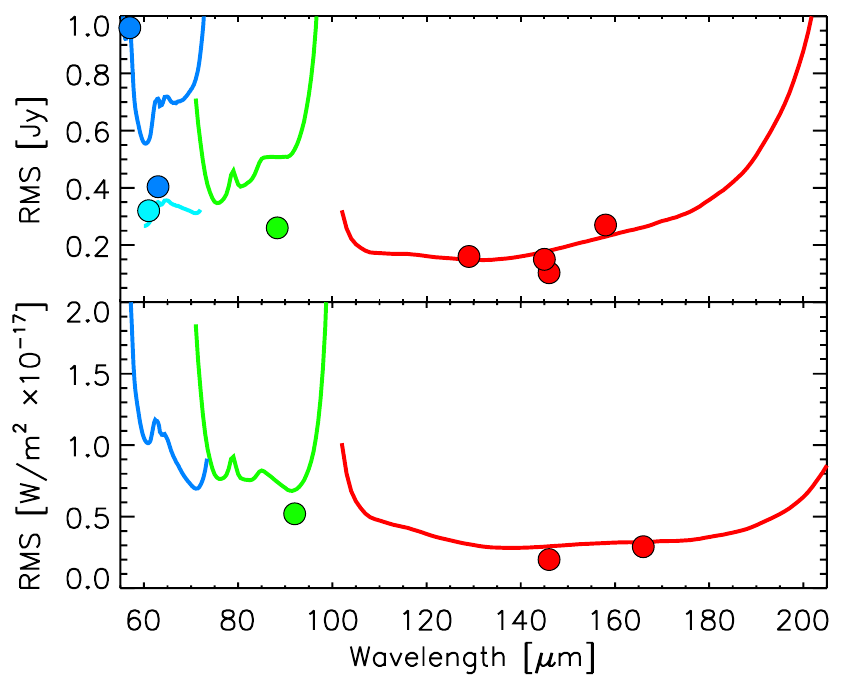}
	\caption{One sigma continuum sensitivity (upper plot) and line
	sensitivity (lower plot) for a number of faint line detections in
	comparison to the HSPOT predictions for a single Nod and single up-down
	scan by the grating, with a total execution time of 400\ldots 440\,s, depending
	on wavelength. The different colours represent the different
	spectral PACS bands and grating orders. Nyquist binning (two bins per
	FWHM) has been used to derive the measured line detection sensitivity
	{\em in each bin}, while the HSPOT prediction refers to total line flux.
	Thus, the actual sensitivity values shown here are very conservative.
              }
         \label{SpecSensitivity}
   \end{figure}

Characterising the sensitivity of the PACS spectrometer resulted in 
rms noise values which are compatible with (and mostly better
than) pre-launch predictions for all spectroscopic observing modes.
Figure~\ref{SpecSensitivity} provides a comparison of HSPOT
sensitivities and several measured values derived from faint line and
continuum observations.

The wavelength dependence of the absolute response of each spectrometer
pixel is characterised by an individual relative spectral response
function. Since this calibration file has been derived from the extended
laboratory black-body measurements, its application to point sources
requires additional diffraction corrections. The required correction
curve is provided in Fig.~\ref{fluxcal}; however, partly extended sources
may consequently show deviating spectral shapes according to their size
and morphology. The precision of the relative spectral response is
affected by spectral leakage (order overlap due to finite steepness of
order sorting filter cut-off edges) from grating order n+1 into grating
order n. At wavelengths of 70--73\,$\mu$m, 98--105\,$\mu$m and 190--220\,$\mu$m 
the next higher grating order wavelengths of 52.5--54.5\,$\mu$m,
65--70\,$\mu$m and 95--110\,$\mu$m do overlap respectively. Continuum shapes
and flux densities in these border ranges are therefore less reliable
than in band centres.

\section{Data analysis and pipeline}

The \emph{Herschel} ground segment (\emph{Herschel} Common Science System - HCSS)
has been implemented using Java technology and written in a common effort by
the \emph{Herschel} Science Centre and the three instrument teams.
One of the HCSS components is the PACS data processing system.

The PACS software was designed for a smooth transition between the different phases of the project, from instrument level tests to routine phase
observations. PACS software supported various operation scenarios: 
\begin{itemize}
\item Pipeline processing:
Pipeline processing is the automatic execution of the standard data processing steps.
(\cite{ewieprec09, schreib09}). Generated products are saved after defined
processing levels).
\item Interactive processing:
The PACS Standard Product Generation (SPG/Pipeline) is designed in a modular way.
Within the \emph{Herschel} Interactive Processing Environment (HIPE) \footnote{HIPE
is a joint development by the Herschel Science Ground Segment Consortium, consisting
of ESA, the NASA Herschel Science Center, and the HIFI, PACS and SPIRE consortia.
See http://herschel.esac.esa.int/DpHipeContributors.shtml}  it is possible
to run the pipeline stepwise, change processing parameter,
add self-defined processing steps, save and restore the intermediate
products, apply different calibrations and inspect the intermediate results.
Various GUI tools support the intermediate and final product inspection.
(\cite{ewieprec09, schreib09})
\end{itemize}
\subsection{Level 0 Product generation}
Level 0 products are complete sets of data as a starting point for scientific data reduction.
Level 0 products may reach a data volume of 750 MB/hour (photometer), or
200 MB/hour (spectrometer).
Depending on the observing mode the data are decompressed and organized in user friendly structures,
namely the Frames class for on-board reduced data (holding basically data cubes) and the Ramps class
containing typical spectrometer ramp data.
Housekeeping (HK) data are stored in tables with converted and raw values (\cite{huygen05}).
Also the full resolution instrument status information (chopper position etc.) is stored as products.
Additional data, like the spacecraft pointing, time correlation, and selected
spacecraft housekeeping information are provided as auxiliary products.
This information is partly merged as status entries into the basic science
products. Finally level 0 contain the calibration data needed for data analysis.
\subsection{Level 0.5 Product generation}
Processing until this level is AOT independent.
Additional information like processing flags and masks (saturation, damaged pixel,
signals affected by chopper and grating transitions)
is added, overviews generated, basic unit conversions applied (digital readouts to Volts/s)
and for the spectrometer the wavelength calibration (inclusive velocity correction) is done.
Also the center of field coordinates are computed for every frame and sky coordinates are
assigned for every pixel.

\subsection{Level 1 Product generation}
The automatic data generation of level 1 products is partly AOT
dependent. Level 1 processing includes the flux calibration and adds
further status information to the product (e.g. chopper angle, masks,
etc.). The resulting product contains the instrument effect free, flux
calibrated data with associated sky coordinates.
These are the biggest products in the processing chain and may reach 3GB/h for photometry and
2GB/h in the case of spectroscopy. The spectroscopy Level 1 product (PacsCube) contains fully calibrated
n$\times$5$\times$5 cubes per pointing/spectral range. The AOT independent steps in
\emph{spectroscopy} are:
\begin{itemize}
 \item Flag glitched signals
 \item Scale signal to standard capacitance
 \item Determine response and dark from calibration blocks
 \item Correct signal non-linearities
 \item  Divide signals by the relative spectral response function
 \item  Divide signals by the response
\end{itemize}
\begin{flushleft}
Chopped AORs:
\end{flushleft}
\begin{itemize}
\item  Subtract off-source signal from on-source signal
\item  Combine the nod positions
\item  Create a PacsCube from PACS Spectrometer Frames
\end{itemize}
\begin{flushleft}
Wavelength switching AORs:
\end{flushleft}
\begin{itemize}
 \item  Subtract the dark
 \item  Subtract the continuum
 \item  Create a PacsCube from PACS Spectrometer Frames
 \item  Compute a differential Frames by pairwise differencing
 \item  Fit the differential frames
 \item  Create a synthetic PacsCube based on the fit
\end{itemize}
\begin{flushleft}
The processing of \emph{photometer} level 1 data is also partly AOT dependent. The resulting product
contains a data cube with flux densities and associated sky coordinates stored as
a PACS Frames class.  Common tasks are:
\end{flushleft}
\begin{itemize}
 \item Process the calibration blocks
 \item Flag permanently damaged pixels
 \item Flag saturated signals
 \item Convert digital readouts to Volts
 \item Correct crosstalk (currently not activated)
 \item Detect and Correct Glitches
 \item Derive UTC from on board time
 \item Convert the digital chopper to angle and angle on sky
 \item Calculate the centre-of-field coordinates for every frame
 \item Responsivity calibration
\end{itemize}
\begin{flushleft}
Chopped point source observations additional tasks:
\end{flushleft}
\begin{itemize}
 \item Flag signals affected by chopper transitions
 \item Mark dither positions
 \item Mark raster positions
 \item Average chopper plateau data
 \item Calculate central pointing per dither and nod position
 \item Calculate sky coordinates for every readout
 \item Compute difference of chopper position readouts
 \item Average dither positions
 \item Compute difference of nod positions
 \item Combine nod data
\end{itemize}
\subsection{Level 2 Product generation}
These data products can be used for scientific analysis. Processing to this
level contains actual images and is highly AOT dependent.
Specific software may be plugged in. For optimal results many of the
processing steps along the route from level 1 to level 2 require
human interaction.
Drivers are both the choice of the right processing parameters as well as
optimising the processing for the scientific goals of the observation.
The result is an Image or Cube product.

\begin{flushleft}
Spectroscopy processing:
\end{flushleft}
The Spectrometer level 2 data product contains noise filtered, regularly
sampled data cubes and a combined cube projected on the WCS.
In the case of wavelength switching an additional cube holds the result
of the fits.

\begin{itemize}
\item Generate a wavelength grid for rebinning
\item  Flag outliers
\item  Rebin on wavelength grid
 \item Project the spaxels onto the WCS combining several 5$\times$5 cubes
\end{itemize}

\begin{flushleft}
Photometry processing:
\end{flushleft}
The PACS Software System offers two options for level 2 generation.
The simple projection -- also used in the automatic pipeline -- and
MADmap processing, which is available only in the interactive
environment as it currently requires significant human interaction.

\begin{flushleft}
Simple projection
\end{flushleft}

\begin{itemize}
\item HighPass filtering.
 The purpose is to remove the 1/f noise. Several methods are still under
 investigation. At the moment the task is using a median filter, which
subtracts a running median from each readout. The filter box size can be set by the user.
 This method is optimised for point sources while low spatial frequencies (extended structures)
 may get suppressed.

\item Projection.
 The task performs a simple coaddition of images, by using a simplified
 version of the drizzle method (\cite{fruchter05})
 It can be applied to scan map observations without any particular
 restrictions.
\end{itemize}

\begin{flushleft}
MADmap
\end{flushleft}
Reconstruction of extended structures requires a different map-making algorithm, which does not remove the low spatial frequencies. PACS adopted the Microwave
Anisotropy Dataset mapper for its optimal map making in the presence of 1/f noise (\cite{cantal10}).
MADmap produces maximum likelihood maps from time-ordered data streams.  The method is based on knowledge of the noise covariance.

In the implementation of MADmap for PACS the noise is represented with an
a-priori determined noise model for each bolometer channel.  In addition,
the correlated (inter-channel) signal is removed prior to map making.

\section{Conclusions}

With the PACS instrument, we have -- for the first time -- introduced large, filled focal plane arrays, as well as integral-field spectroscopy with diffraction-limited resolution, in the far infrared.  Without any prior demonstration, several major, technological developments have found their first application with PACS in space.  The success of this path -- documented by the results in this volume -- should encourage our community to defend this more ``pioneering'' approach against trends towards an ``industrial'' (i.e., minimal-risk) approach, which will not allow us to take advantage of the latest, experimental developments, on which our scientific progress often depends.

\begin{acknowledgements}

PACS has been developed by a consortium of institutes led by MPE  (Germany) and including UVIE (Austria); KU Leuven, CSL, IMEC (Belgium); CEA, LAM (France); MPIA (Germany); INAF-IFSI/OAA/OAP/OAT, LENS, SISSA  (Italy); IAC (Spain). This development has been supported by the funding agencies BMVIT (Austria), ESA-PRODEX (Belgium), CEA/CNES (France),
DLR (Germany), ASI/INAF (Italy), and CICYT/MCYT (Spain).

AP would like to thank the entire PACS consortium, including industrial partners, for following on this crazy adventure, and T. Passvogel and the Project Team at ESA for tolerating us.  F. Marliani's (ESTEC) support in fighting a crucial battle shall never be forgotten.

\end{acknowledgements}


\begin{thebibliography}{}

\bibitem[Ade et al. 2006]{ade06} Ade, P.A.R., Pisano, G., Tucker, C., \& Weaver, S. 2006,
     Proc. SPIE, 6275, 62750U
     
\bibitem[Agnese et al. 1999]{Agnese99} Agnese, P., Buzzi, C., Rey, P., Rodriguez, L., \& Tissot, J.-L. 1999, Proc. SPIE, 3698, 284

\bibitem[Agnese et al. 2003]{Agnese03} Agnese, P., et al 2003, Proc. SPIE, 4855, 108

\bibitem[Billot et al. 2007]{Billot07} Billot, N., et al 2007, Proc. SPIE, 6265, 9B

\bibitem[Cantalupo and Hook, 2010]{cantal10} Cantalupo, C. M., Borrill, J. D., Jaffe, A. H., Kisner, T. S., \& Stompor, R. 2010, ApJS, 187, 212

\bibitem[Dehaes et al. 2010]{Dehaes10} Dehaes, S., Bauwens, E., Decin, L., et al. 2010, A\&A (submitted) 

\bibitem[Duband et al. 2008]{duband08} Duband. L., Clerc, L., Ercolani, E., Guillemet, L., \& Vallcorba, R. 2008, Cryogenics 48, 95

\bibitem[Fischer et al. 2004]{fischer04} Fischer, J., Klaassen, T., Hovenier, N., Jakob, G., Poglitsch, A., \& Stenberg, O. 2004, App. Opt., 43, 3765 

\bibitem[Fletcher at al. 2010]{Fletch10} Fletcher, L. N., Drossart, P., Burgdorf, M., et al. 2010,
A\&A, 514, A17

\bibitem[Fruchter and Hook, 2002]{fruchter05} Fruchter, A. S., \& Hook, R. N. 2002, PASP, 114, 144.

\bibitem[Huygen et al. 2006]{huygen05} Huygen, R., Vandenbussche, B., \& Wieprecht, E. 2006,  ASP Conference Series, 351, 220

\bibitem[Inguscio et al. 1986] {inguscio86} Inguscio, M., Moruzzi, G., Evenson, K., Jennings, \& D. 1986, J. Appl. Phys., 60, 161

\bibitem[Katterloher et al. 2006]{Katterloher06} Katterloher, R., Barl, L., Poglitsch, A., Royer, P., \&
Stegmaier, J. 2006, Proc. SPIE, 6275, 627515

\bibitem[Kraft et al. 2000]{Kraft00} Kraft, S., Frenzl, O., Charlier, O.,  Cronje, T.,
	 	 Katterloher, R.O.,  Rosenthal, D., Gr\"ozinger, U., \& Beeman, J.W. 2000,
		 Proc. SPIE, 4013, 233

\bibitem[Kraft et al. 2001]{Kraft01} Kraft, S., et al. 2001, Proc. SPIE, 4540, 374

\bibitem[Krause et al. 2006]{krause06} Krause, O., Lemke, R., Hofferbert, R., B\"ohm, A., Klaas, U., 
         Katzer, J., H\"oller, F., \& Salvasohn, M. 2006, Proc. SPIE, 6273, 627325

\bibitem[Looney et al. 2003]{Looney03} Looney, L.W., Raab, W., Poglitsch, A., Geis, N. 2003, ApJ, 597, 628

\bibitem[Merken et al. 2004]{Merken04} Merken, P., Creten, Y., Putzeys, J., Souverijns, T., \& Van Hoof, C. 2004, Proc. SPIE, 5498, 622

\bibitem[M\"uller \& Lagerros 1998]{MulLag98} M\"uller, T. G., Lagerros, J. S. V. 1998, A\&A, 338, 340

\bibitem[M\"uller \& Lagerros 2002]{MulLag02} M\"uller, T. G., Lagerros, J. S. V. 2002, A\&A, 381, 324

\bibitem[Ottensamer \& Kerschbaum 2008]{Ottensamer08} Ottensamer, R., \& Kerschbaum, F. 2008, Proc. SPIE, 7019, 70191B

\bibitem[PACS Observer's Manual]{pacs10} PACS Observer's Manual, 2010, http://herschel.esac.esa.int/Docs/PACS/pdf/pacs\_om.pdf

\bibitem[Pilbratt et al. 2010]{pilbratt10} Pilbratt, G., et al. 2010, A\&A this issue

\bibitem[Poglitsch et al. 2003]{PogDet03} Poglitsch, A., et al. 2003, Proc. SPIE, 4855, 115

\bibitem[Poglitsch et al. 2006]{alpog06} Poglitsch, A., et al. 2006,
      Proc. SPIE, 6265, 62650B

\bibitem[Poglitsch et al. 2008]{alpog08} Poglitsch, A., et al. 2008,
      Proc. SPIE, 7015, 701005

\bibitem[Poglitsch \& Altieri 2009]{alpog09} Poglitsch, A., \& Altieri, B. 2009, EAS Publications Series,
      34, 43
      
\bibitem[Renotte et al. 1999]{Renotte99} Renotte, E., Gillis, J.-M., Jamar, C.A., Laport, P., Sal\'ee, T.,   \& Crehay, S. 1999, Proc. SPIE, 3759, 189

\bibitem[Schreiber et al. 2009]{schreib09} Schreiber, J., Wieprecht, E., de Jong, J., Wetzstein, M., Jacobson, J., et al. 2009, ASP Conference Series, 411, 478

\bibitem[Schulz et al. 2002]{Schulz02} Schulz, B., Huth, S., Laureijs, R. J., et al. 2002, A\&A, 381, 1110

\bibitem[Shirahata et al. 2009]{Shira09} Shirahata, M., Matsuura, S., Hasegawa, S., et al. 2009,
  PASJ, 61, 737

\bibitem[Simoens et al. 2004]{Simoens04} Simoens, F., et al 2004, Proc. SPIE, 5498, 177

\bibitem[Stegmaier et al. 2008]{Stegmaier08} Stegmaier, J., Birkmann, S., Gr\"ozinger, U., Krause, O., \& Lemke, D. 2008, Proc. SPIE, 7010, 701009

\bibitem[Wieprecht et al. 2009]{ewieprec09} Wieprecht, E., Schreiber, J., de Jong, J., Jacobson, J., Liu, C., et al. 2009, ASP Conference Series, 411, 531

  \end{thebibliography}
\end{document}